\documentclass[reprint,amsmath,amssymb,aps,prb,superscriptaddress]{revtex4-1}
\usepackage{array}
\usepackage{bm}
\usepackage{color}
\usepackage{float}
\usepackage{graphicx}
\usepackage{hyperref}
\hypersetup{colorlinks=true,allcolors=blue}
\usepackage{natbib}
\usepackage{newtxtext}
\usepackage{newtxmath}
\usepackage[caption=false]{subfig}

\begin{document}

\title{Theoretical study of quantum spin liquids in $S=1/2$ hyper-hyperkagome magnets: \\ classification, heat capacity, and dynamical spin structure factor}

\author{Li Ern Chern}
\affiliation{Department of Physics, University of Toronto, Toronto, Ontario M5S 1A7, Canada}

\author{Yong Baek Kim}
\affiliation{Department of Physics, University of Toronto, Toronto, Ontario M5S 1A7, Canada}


\begin{abstract}
Recent experiments suggest a quantum spin liquid ground state in the material PbCuTe$_2$O$_6$, where $S=1/2$ moments are coupled by antiferromagnetic Heisenberg interactions into a three dimensional structure of corner sharing triangles dubbed the hyper-hyperkagome lattice. It exhibits a richer connectivity, and thus likely a stronger geometric frustration, than the relatively well studied hyperkagome lattice. Here, we investigate the possible quantum spin liquids in the $S=1/2$ hyper-hyperkagome magnet using the complex fermion mean field theory. Extending the results of a previous projective symmetry group analysis, we identify only two $\mathbb{Z}_2$ spin liquids and a $U(1)$ spin liquid that are compatible with the hyper-hyperkagome structure. The $U(1)$ spin liquid has a spinon Fermi surface. For the $\mathbb{Z}_2$ spin liquids, one has a small excitation gap, while the other is gapless and proximate to the $U(1)$ spin liquid. We show that the gapped and gapless spin liquids can in principle be distinguished by heat capacity measurements. Moreover, we calculate the dynamical spin structure factors of all three spin liquids and find that they highly resemble the inelastic neutron scattering spectra of PbCuTe$_2$O$_6$. Implications of our work to the experiments, as well as its relations to the existing theoretical studies, are discussed.
\end{abstract}


\maketitle


\section{Introduction}

Frustrated $S=1/2$ magnets have long been considered as promising platforms for the discovery of emergent ground states with unusual physical properties. Due to the presence of competing interactions and quantum fluctuations, the spins may evade magnetic ordering down to zero temperature, forming a quantum spin liquid phase\cite{nature08917,Savary_2016,RevModPhys.89.025003}. One way to achieve frustration is to have spins arranged in triangular motifs and coupled by antiferromagnetic interaction, which is known as geometric frustration\cite{annurev.ms.24.080194.002321}. Examples are triangular and kagome systems in two dimensions, and pyrochlore and hyperkagome systems in three dimensions. A number of quantum spin liquid candidate materials that realize these structures and a dominant nearest neighbor antiferromagnetic Heisenberg exchange have been identified, which include the organic compound $\kappa$-(BEDT-TTF)$_2$Cu$_2$(CN)$_3$ (triangular)\cite{PhysRevLett.91.107001,PhysRevB.73.140407,nphys942,nphys1134}, herbertsmithite (kagome)\cite{PhysRevLett.98.077204,PhysRevLett.98.107204,PhysRevLett.100.077203,nature11659,Fu655}, and Na$_4$Ir$_3$O$_8$ (hyperkagome)\cite{PhysRevLett.99.137207,PhysRevB.88.220413,PhysRevLett.115.047201}.

Among these lattices with triangular motifs, hyperkagome has arguably the most complicated structure. The $S=1/2$ hyperkagome magnet Na$_4$Ir$_3$O$_8$ has attracted numerous theoretical efforts to identify the possible spin liquid and magnetically ordered states\cite{PhysRevLett.99.037201,PhysRevLett.100.227201,PhysRevB.78.094403,PhysRevLett.101.197201,PhysRevLett.101.197202,PhysRevB.85.104406,PhysRevB.87.165120,PhysRevB.93.094419,PhysRevB.94.064416,PhysRevB.94.235138,PhysRevB.95.054404}. It belongs to the space group P$4_132$ with 12 sites per unit cell\cite{PhysRevLett.99.137207}. The nearest neighbor bonds form a three-dimensional network of corner sharing triangles, where each site is shared by two triangles. Yet, recent experiments\cite{PhysRevB.90.035141,PhysRevLett.116.107203,s41467-020-15594-1} on the $S=1/2$ magnet PbCuTe$_2$O$_6$ reveal an even more elaborate structure. Belonging to the same space group P$4_132$, it can be viewed as a distorted version of Na$_4$Ir$_3$O$_8$, which results in interactions beyond the first nearest neighbors and thus an enhanced connectivity between the sites. In particular, the first and second nearest neighbor bonds also form corner sharing triangles in three dimensions, but each site now participates in three triangles. Such a structure is dubbed the hyper-hyperkagome lattice\cite{s41467-020-15594-1}.

Multiple measurements on PbCuTe$_2$O$_6$ have pointed to a quantum spin liquid ground state. Magnetic susceptibility follows a Curie Weiss behavior and shows no sign of long range order down to $2$ K\cite{PhysRevB.90.035141}. The Curie Weiss temperature is found to be $-22$ K, which indicates the overall energy scale and antiferromagnetic nature of the spin interactions. Heat capacity shows a broad peak roughly at $1$ K, which is unlikely a result of magnetic ordering or spin freezing\cite{PhysRevB.90.035141}. Such a broad peak in heat capacity is similarly observed in the quantum spin liquid candidate Na$_4$Ir$_3$O$_8$. Muon spin relaxation does not detect any signal of magnetic order down to $0.02$ K, but reveals persistent slow spin dynamics below $1$ K, which is also confirmed by nuclear magnetic resonance\cite{PhysRevLett.116.107203}. More recently, inelastic neutron scattering (INS) finds a dispersionless, diffusive continuum of signals suggestive of fractionalized excitations in a quantum spin liquid\cite{s41467-020-15594-1}.

Motivated by these experiments, we theoretically investigate the possible quantum spin liquids in the $S=1/2$ hyper-hyperkagome magnet. We consider the antiferromagnetic Heisenberg model of PbCuTe$_2$O$_6$ up to the second nearest neighbor interaction, with $J_1=J_2$ based on the most recent density functional theory (DFT) estimation\cite{s41467-020-15594-1}. We use the complex fermion mean field theory\cite{PhysRevB.44.2664,PhysRevB.49.5200,PhysRevB.65.165113,PhysRevB.83.224413,PhysRevB.95.054404} which expresses the spin Hamiltonian in terms of hopping and pairing of spinons. Previously, the projective symmetry group (PSG) analysis\cite{PhysRevB.65.165113,PhysRevB.83.224413} had been applied to classify the possible quantum spin liquids in the $S=1/2$ hyperkagome magnet\cite{PhysRevB.95.054404,PhysRevB.101.054408}. Since the PSG analysis depends only on the symmetry but not the microscopic model, we can extend the results in Ref.~\onlinecite{PhysRevB.95.054404} rather straightforwardly to the $S=1/2$ hyper-hyperkagome magnet. We show that, among the five $\mathbb{Z}_2$ spin liquids (two $U(1)$ spin liquids) that respect the P$4_132$ space group and time reversal symmetry, only two (one) are physical on the hyper-hyperkagome structure.

\begin{table*}
\caption{\label{windmill} The coordinates of the 12 sublattices, labeled by $s$ (first column), of the distorted windmill lattice (second column), the hyperkagome lattice (third column) and the hyper-hyperkagome lattice (fourth column). The most generic forms of the $12d$ special coordinates of the P$4_132$ space group, given here as the 12 sublattices of the distorted windmill lattice, are parametrized by a continuous variable $y$. The hyperkagome and hyper-hyperkagome lattices take the specific values $y=-1/8$ and $y=-0.2258$ respectively. Each entry of the sublattice coordinates in the distorted windmill lattice is defined up to modulo $1$, i.e.~a unit cell translation in the respective cubic direction. For example, take $s=1$ and $y=-1/8$, the $z$-coordinate is $1-y=1+1/8=1/8 \, (\mathrm{mod} \, 1)$.}
\begin{ruledtabular}
\begin{tabular}{cccc}
$s$ & distorted windmill & hyperkagome ($y=-1/8$) & hyper-hyperkagome ($y=-0.2258$) \\ \hline
1 & $(3/4+y,3/8,1-y)$ & $(5/8,3/8,1/8)$ & $(0.5242,0.375,0.2258)$ \\
2 & $(1/2+y,1/4-y,7/8)$ & $(3/8,3/8,7/8)$ & $(0.2742,0.4758,0.875)$ \\
3 & $(5/8,1/2-y,3/4-y)$ & $(5/8,5/8,7/8)$ & $(0.625,0.7258,0.9758)$ \\
4 & $(1/2-y,3/4-y,5/8)$ & $(5/8,7/8,5/8)$ & $(0.7258,0.9758,0.625)$ \\
5 & $(3/4-y,5/8,1/2-y)$ & $(7/8,5/8,5/8)$ & $(0.9758,0.625,0.7258)$ \\
6 & $(7/8,1/2+y,1/4-y)$ & $(7/8,3/8,3/8)$ & $(0.875,0.2742,0.4758)$ \\
7 & $(1-y,3/4+y,3/8)$ & $(1/8,5/8,3/8)$ & $(0.2258,0.5242,0.375)$ \\
8 & $(1/4-y,7/8,1/2+y)$ & $(3/8,7/8,3/8)$ & $(0.4758,0.875,0.2742)$ \\
9 & $(3/8,1-y,3/4+y)$ & $(3/8,1/8,5/8)$ & $(0.375,0.2258,0.5242)$ \\
10 & $(1/4+y,1/8,y)$ & $(1/8,1/8,7/8)$ & $(0.0242,0.125,0.7742)$ \\
11 & $(1/8,y,1/4+y)$ & $(1/8,7/8,1/8)$ & $(0.125,0.7742,0.0242)$ \\
12 & $(y,1/4+y,1/8)$ & $(7/8,1/8,1/8)$ & $(0.7742,0.0242,0.125)$ \\
\end{tabular}
\end{ruledtabular}
\end{table*}

We then solve the mean field self consistent equations for these spin liquid states. We find that the $U(1)$ spin liquid is gapless with a Fermi surface of spinons. One of the $\mathbb{Z}_2$ spin liquids has a small excitation gap of the order $0.01 J_1$, while the other is gapless and proximate to the $U(1)$ spin liquid. All of them appear to be consistent with the experimental observations that a gap, if exists, should be less than $0.15$ meV\cite{s41467-020-15594-1} or $0.45$ K\cite{PhysRevLett.116.107203}, assuming $J_1 \approx 1 \, \mathrm{meV}$\cite{s41467-020-15594-1}. At the mean field level, the gapped $\mathbb{Z}_2$ spin liquid is the lowest energy state. However, we show that all three states can give rise to dynamical spin structure factors very similar to the observed INS spectra. This result provides further support for a quantum spin liquid ground state in PbCuTe$_2$O$_6$, which is likely to be one of the three spin liquids considered in this work. We further show that the two $\mathbb{Z}_2$ spin liquids can in principle be distinguished by heat capacity in the low temperature limit, where the heat capacity coefficient $C/T$ vanishes (remains finite) for the gapped (gapless) state.

The rest of the paper is organized as follows. In Sec.~\ref{modelsection}, we describe the spin model under study, as well as the structure and the symmetry of the hyper-hyperkagome lattice. In Sec.~\ref{meanfieldsection}, we introduce the complex fermion mean field theory via the parton construction and extend the PSG analysis in Ref.~\onlinecite{PhysRevB.95.054404} to the $S=1/2$ hyper-hyperkagome magnet. In Sec.~\ref{bandsection}, we present the mean field self consistent solutions for the physical $\mathbb{Z}_2$ and $U(1)$ spin liquids and examine the resulting spinon spectra. In Secs.~\ref{heatsection} and \ref{dynamicsection}, we calculate the heat capacities and the dynamical spin structure factors of these spin liquids. We compare the latter to the INS data in Ref.~\onlinecite{s41467-020-15594-1}. In Sec.~\ref{discussion}, we summarize our work and discuss its relation to existing theoretical studies on PbCuTe$_2$O$_6$.

\section{\label{modelsection}Model and Structure}

\begin{figure}
\includegraphics[scale=0.4]{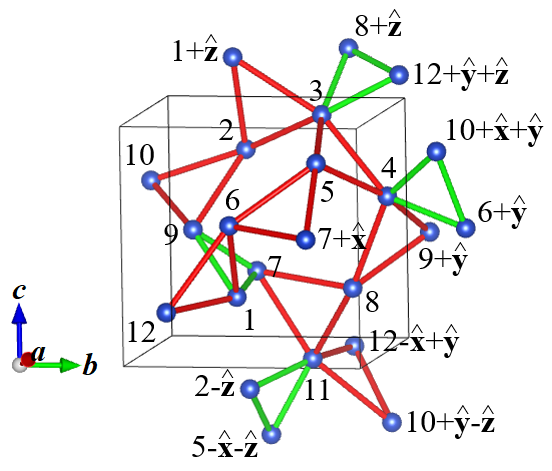}
\caption{\label{lattice}The hyper-hyperkagome lattice. Blue spheres represent Cu$^{2+}$ ions which carry $S=1/2$ moments. Green and red bonds represent the first and second nearest neighbor bonds, respectively. The numbers $1,\ldots,12$ are the sublattice indices. $\hat{\mathbf{x}}$, $\hat{\mathbf{y}}$, and $\hat{\mathbf{z}}$ are the unit cell translations along the $x$, $y$, and $z$ directions, respectively. Bonds that are equivalent up to lattice translations are only plotted once.}
\end{figure}

Magnetism in PbCuTe$_2$O$_6$ is due to Cu$^{2+}$ ions carrying $S=1/2$. PbCuTe$_2$O$_6$ belongs to the cubic space group P$4_132$; all the Cu$^{2+}$ sites are crystallographically equivalent\cite{PhysRevB.90.035141,PhysRevLett.116.107203,s41467-020-15594-1}. According to a recent density functional theory calculations\cite{s41467-020-15594-1}, the $S=1/2$ moments are coupled by antiferromagnetic Heisenberg interactions up to the fourth nearest neighbor. The ratio of these interactions are estimated as $J_1:J_2:J_3:J_4 \approx 1:1:0.5:0.1$. It is argued that the magnetic frustration arises from the dominant $J_1$ and $J_2$ interactions with approximately equal strength, which leads to an infinite classical ground state degeneracy\cite{s41467-020-15594-1}. Therefore, we consider the $J_1$-$J_2$ model with $J_1=J_2$ in this work for simplicity,
\begin{equation} \label{J1J2model}
H = J_1 \sum_{\langle ij \rangle} \mathbf{S}_i \cdot \mathbf{S}_j + J_2 \sum_{\langle \langle ij \rangle \rangle} \mathbf{S}_i \cdot \mathbf{S}_j .
\end{equation}
This results in a three dimensional structure of corner sharing triangles, in which each site is shared by three triangles. Such a structure is dubbed the hyper-hyperkagome lattice\cite{s41467-020-15594-1}, due to its similarity to the hyperkagome lattice\cite{PhysRevLett.99.137207,PhysRevB.95.054404,PhysRevB.101.054408}, in which each site is a common vertex of two triangles--one less than the former. Indeed, if we consider only the $J_2$ interaction in the hyper-hyperkagome lattice, the connectivity of the sites is equivalent to the nearest neighbor hyperkagome model. On the other hand, the $J_1$ bonds in the hyper-hyperkagome lattice form isolated triangles. See Fig.~\ref{lattice} for an illustration of the hyper-hyperkagome structure.

Both hyperkagome and hyper-hyperkagome lattices belong to the same space group P$4_132$. Both have 12 sublattices per unit cell, but different sets of positions for these sublattices. We refer to the distorted windmill lattice\cite{PhysRevB.78.014404} for the generic coordinates of the $12d$ special positions of the P$4_132$ space group\cite{PhysRevB.101.054408}, at which the 12 sublattices are located. The generic coordinates are parametrized by a continuous variable $y$, see Table~\ref{windmill}. The hyperkagome lattice corresponds to $y=-1/8=-0.125$\cite{PhysRevB.95.054404}, while the hyper-hyperkagome lattice corresponds to $y=-0.2258$ according to the structural parameters provided by Ref.~\onlinecite{s41467-020-15594-1}. This allows us to label the 12 sublattices independent of the precise value of $y$, and examine how they evolve when $y$ is being tuned from one value to another. For example, when $y$ is changed from $-0.125$ to $-0.2258$, the nearest neighbor bonds in the hyperkagome lattice becomes the second nearest neighbor bonds in the hyper-hyperkagome lattice. Realizing such a unified description is helpful as we can extend the previously established classification of symmetric spin liquids in Ref.~\onlinecite{PhysRevB.95.054404} on the hyperkagome lattice to the hyper-hyperkagome lattice, which is described in the next section.

As discussed in Ref.~\onlinecite{PhysRevB.95.054404}, the P$4_132$ space group is generated by (i) a three-fold rotation $C_3$ about the $(1,1,1)$ axis, (ii) a two-fold rotation $C_2$ about the $(3/8,3/4-x_2,x_2)$ axis, and (iii) a four-fold screw $S_4$, which consists of a $\pi/2$ rotation about the $(x_1,-1/4,1/2)$ axis followed by a fractional translation of $(1/4,0,0)$. Their actions on a generic point with coordinates $(x,y,z)$ are
\begin{subequations}
\begin{align}
& C_3: (x,y,z) \longrightarrow \left( z,x,y \right), \label{C3operation} \\
& C_2: (x,y,z) \longrightarrow \left( \frac{3}{4}-x,\frac{3}{4}-z,\frac{3}{4}-y \right), \label{C2operation} \\
& S_4: (x,y,z) \longrightarrow \left( \frac{1}{4}+x,\frac{1}{4}-z,\frac{3}{4}+y \right) . \label{S4operation}
\end{align}
\end{subequations}
Notice that $S_4$ is a nonsymmorphic symmetry. Translation along the $x$ direction, $T_1: (x,y,z) \longrightarrow (x+1,y,z)$, is generated by $S_4$ via $T_1=(S_4)^4$. Translations along the $y$ and $z$ directions, $T_2$ and $T_3$, are related to $T_1$ by $C_3$. In practice, it is convenient to express the coordinates of the sites on the (hyper-)hyperkagome lattice as $(x,y,z;s)$, where $(x,y,z) \in \mathbb{Z} \times \mathbb{Z} \times \mathbb{Z}$ labels the unit cell and $s \in \lbrace 1, \ldots, 12 \rbrace$ labels the sublattice. The action of the operators \eqref{C3operation}-\eqref{S4operation} on the lattice sites can be found in Table \ref{rotationscrew} in Appendix \ref{ansatzappendix}.

\section{\label{meanfieldsection}Complex Fermion Mean Field Theory and Classification of Symmetric Spin Liquids}

\subsection{Parton Construction and Projective Symmetry Group}

To explore the quantum spin liquid phases with deconfined spinons in the $S=1/2$ hyper-hyperkagome magnet, we first apply the complex fermion mean field theory\cite{PhysRevB.44.2664,PhysRevB.49.5200,PhysRevB.65.165113,PhysRevB.83.224413,PhysRevB.95.054404} to the spin Hamiltonian \eqref{J1J2model}. The spin operator is represented in terms of fermionic spinons (partons)
\begin{equation} \label{partonrepresentation}
\mathbf{S}_i = \frac{1}{2} \sum_{\alpha \beta} f_{i \alpha}^\dagger \vec{\sigma}_{\alpha \beta} f_{i \beta} .
\end{equation}
The Hamiltonian is now quartic in the fermionic spinons. For an antiferromagnetic Heisenberg exchange, the spin interaction can be rewritten in the form
\begin{subequations}
\begin{align}
\mathbf{S}_i \cdot \mathbf{S}_j &= - \frac{1}{4} \left( \hat{\chi}_{ij}^\dagger \hat{\chi}_{ij} + \hat{\Delta}_{ij}^\dagger \hat{\Delta}_{ij} \right) , \label{singletchannel} \\
\hat{\chi}_{ij} &= f_{i \uparrow}^\dagger f_{j \uparrow} + f_{i \downarrow}^\dagger f_{j \downarrow} , \label{singlethopping} \\
\hat{\Delta}_{ij} &= f_{i \uparrow} f_{j \downarrow} - f_{i \downarrow} f_{j \uparrow} . \label{singletpairing}
\end{align}
\end{subequations}
\eqref{singlethopping} and \eqref{singletpairing} are known as the \textit{singlet hopping} and the \textit{singlet pairing} of spinons, respectively. A mean field decoupling of \eqref{singletchannel} then results in a quadratic and thus solvable Hamiltonian,
\begin{equation} \label{meanfieldhamiltonian}
\begin{aligned}[b]
H^\mathrm{MF} &= - \sum_{ij} \frac{J_{ij}}{4} \left[ \left( \chi_{ij}^* \hat{\chi}_{ij} + \chi_{ij} \hat{\chi}_{ij}^\dagger - \lvert \chi_{ij} \rvert^2 \right) \right. \\
& \left. + \left( \Delta_{ij}^* \hat{\Delta}_{ij} + \Delta_{ij} \hat{\Delta}_{ij}^\dagger - \lvert \Delta_{ij} \rvert^2 \right) \right] \\
&+ \sum_i \lambda_i^{(3)} \left( f_{i \uparrow}^\dagger f_{i \uparrow} + f_{i \downarrow}^\dagger f_{i \downarrow} - 1 \right) \\ 
&+ \sum_i \left[ (\lambda_i^{(1)}+i\lambda_i^{(2)}) f_{i \downarrow} f_{i \uparrow} + (\lambda_i^{(1)}-i\lambda_i^{(2)}) f_{i \uparrow}^\dagger f_{i \downarrow}^\dagger \right] .
\end{aligned}
\end{equation}
$\chi_{ij}$ and $\Delta_{ij}$ (without hats) are variational parameters, not operators. On-site Lagrange multipliers $\lambda_i^{(1)},\lambda_i^{(2)},\lambda_i^{(3)} \in \mathbb{R}$ are introduced to enforce the single occupancy constraint (i.e.~one spinon per site), as the parton representation \eqref{partonrepresentation} allows zero and double occupancies which are unphysical. Minimizing the mean field energy with respect to the variational parameters yields the self consistent equations $\chi_{ij} = \langle \hat{\chi}_{ij} \rangle$ and $\Delta_{ij} = \langle \hat{\Delta}_{ij} \rangle$, which are usually solved in the momentum space through the Fourier transform,
\begin{equation} \label{fourier}
f_{\mathbf{k},s,\alpha} = \frac{1}{\sqrt{N}} \sum_\mathbf{R} f_{\mathbf{R},s,\alpha} e^{-i \mathbf{k} \cdot \mathbf{R}},
\end{equation}
where $\mathbf{R}$, $s$, and $\alpha$ denotes the unit cell coordinates, sublattice index, and the spin flavor, respectively.

The parton representation of spin \eqref{partonrepresentation} introduces an $SU(2)$ gauge redundancy\cite{PhysRevB.38.745,PhysRevB.38.2926}, which is most apparent when written in the form\cite{PhysRevB.95.054404}
\begin{equation}
S_i^\mu = \frac{1}{4} \mathrm{Tr} \left( \Psi_i^\dagger \sigma^\mu \Psi_i \right), \, \Psi_i=\begin{pmatrix} f_{i \uparrow} & f_{i \downarrow}^\dagger \\ f_{i \downarrow} & -f_{i \uparrow}^\dagger \end{pmatrix} . 
\end{equation}
The physical spin $\mathbf{S}_i$ is invariant under the transformation $\Psi_i \longrightarrow \Psi_i G_i, G_i \in SU(2)$, which also preserves the fermionic anticommutation relation. As a consequence, the symmetries of the original Hamiltonian are realized projectively at the mean field level. That is, the mean field Hamiltonian
\begin{equation}
\begin{aligned}[b]
& H^\mathrm{MF} = \sum_{ij} \mathrm{Tr} \left( \Psi_i u_{ij} \Psi_j^\dagger \right) , \, u_{ij}= \frac{J_{ij}}{4} \begin{pmatrix} \chi_{ij} & -\Delta_{ij}^* \\ -\Delta_{ij} & -\chi_{ij}^* \end{pmatrix}, \\
& u_{ii} = - \frac{1}{2} \begin{pmatrix} \lambda_i^{(3)} & \lambda_i^{(1)}+i\lambda_i^{(2)} \\ \lambda_i^{(1)}-i\lambda_i^{(2)} & -\lambda_i^{(3)} \end{pmatrix} ,
\end{aligned}
\end{equation}
is invariant under a symmetry operation $X$ only up to a gauge transformation $G_X$,
which implies that the mean field ansatz satisfies
\begin{equation} \label{ansatzsymmetry}
u_{X(i)X(j)} = G_X(X(i)) u_{ij} G_X(X(j))^\dagger .
\end{equation}
Ref.~\onlinecite{PhysRevB.65.165113} points out that different spin liquids obeying the same set of symmetries $\lbrace X\rbrace$ can be classified according to different sets of gauge transformations $\lbrace G_X \rbrace$. This is the essence of projective symmetry group (PSG) analysis. PSG consists of compound operators of the form $G_X X$. Pure gauge transformations $G_I$ (associated with the identity operator $X=I$) that leave the mean field Hamiltonian invariant forms a subgroup of PSG called the invariant gauge group (IGG). When both hopping and pairing terms are present in the Hamiltonian, as in \eqref{meanfieldhamiltonian}, the IGG is $\lbrace +1, -1 \rbrace$, and the resulting spin liquid is called a $\mathbb{Z}_2$ spin liquid. On the other hand, if the Hamiltonian only contain hopping terms, the IGG is $\lbrace e^{i \theta \tau_3} \vert 0 \leq \theta < 2 \pi \rbrace$, and the resulting spin liquid is called a $U(1)$ spin liquid.

Besides the space group, one typically considers the time reversal symmetry, which is implemented projectively as\cite{PhysRevB.65.165113,PhysRevB.83.224413,PhysRevB.95.054404}
\begin{equation} \label{ansatztimereversal}
u_{ij} = - G_\mathcal{T}(i) u_{ij} G_\mathcal{T} (j)^\dagger .
\end{equation}

The PSG classification of symmetric $\mathbb{Z}_2$ and $U(1)$ spin liquids has been performed on the $S=1/2$ hyperkagome magnet\cite{PhysRevB.95.054404}. Although the lattice structure is complicated, at the end there are only five possible $\mathbb{Z}_2$ spin liquids and two possible $U(1)$ spin liquids that respect the P$4_132$ space group and the time reversal symmetry. Since the PSG analysis depends only on the symmetry but not the microscopic spin interactions, the results of the classification in Ref.~\onlinecite{PhysRevB.95.054404} can be applied to the $S=1/2$ hyper-hyperkagome magnet.

We summarize the solution of the PSG analysis in Ref.~\onlinecite{PhysRevB.95.054404} here. In all cases, the gauge transformations associated with translations $T_{i=1,2,3}$ and $C_3$ are trivial, $G_{T_i}(x,y,z;s)=1$ and $G_{C_3}(x,y,z;s)=1$. The $\mathbb{Z}_2$ or $U(1)$ spin liquids are distinguished by the gauge transformations of the remaining symmetry operators $C_2$, $S_4$, and $\mathcal{T}$. One can choose a gauge in which they are site independent, i.e.~$G_X(x,y,z;s)=g_X$, and $g_{C_2}=g_{S_4}$. The forms of $g_{C_2}$, $g_{S_4}$, and $g_\mathcal{T}$ are shown in Table \ref{hyperkagomepsg}. Throughout this paper, we use the same labels for the $\mathbb{Z}_2$ and $U(1)$ spin liquids as in Ref.~\onlinecite{PhysRevB.95.054404}.

\begin{table}
\caption{\label{hyperkagomepsg} The gauge transformations associated with the symmetry operations $C_2$, $S_4$, and $\mathcal{T}$ for the five possible $\mathbb{Z}_2$ spin liquids and the two possible $U(1)$ spin liquids, from the PSG analysis in Ref.~\onlinecite{PhysRevB.95.054404}. In all cases, the gauge transformations associated with translations and $C_3$ are trivial. When applied to the hyper-hyperkagome lattice, some PSG are unphysical as they yield vanishing mean field ansatzes where they should be finite.}
\begin{ruledtabular}
\begin{tabular}{ccccc}
label & IGG & $g_{C_2}=g_{S_4}$ & $g_\mathcal{T}$ & physical? \\ \hline
1(a) & $\mathbb{Z}_2$ & $1$ & $1$ & no \\
1(b) & $\mathbb{Z}_2$ & $i \tau_3$ & $1$ & no \\
2(a) & $\mathbb{Z}_2$ & $1$ & $i \tau_2$ & yes \\
2(b) & $\mathbb{Z}_2$ & $i \tau_2$ & $i \tau_2$ & no \\
2(c) & $\mathbb{Z}_2$ & $i \tau_3$ & $i \tau_2$ & yes \\
$U1^0$ & $U(1)$ & $1$ & $i \tau_1$ & yes \\
$U1^1$ & $U(1)$ & $i \tau_1$ & $i \tau_1$ & no
\end{tabular}
\end{ruledtabular}
\end{table}

\subsection{Identification of Physical Spin Liquid States}

Now we can identify the physical spin liquid states by analyzing the symmetry constraints on the mean field ansatz. The bond parameters $\chi_{ij}$ and $\Delta_{ij}$, as well as the Lagrange multipliers $\lambda_i$, are constrained by the PSG through \eqref{ansatzsymmetry} and \eqref{ansatztimereversal}. Not all of the seven spin liquids in Table \ref{hyperkagomepsg} are physical, however, as some leads to vanishing bond parameters for the first and/or second nearest neighbors. For 1(a) and 1(b), $g_\mathcal{T}=1$ is trivial, and it is easy to see from \eqref{ansatztimereversal} that time reversal symmetry constrains $u_{ij}=0$ for any pair of sites $i$ and $j$. For 2(b) and $U1^1$, one can show that $u_{ij}=0$ for the first nearest neighbors. As a result, 2(a), 2(c), and $U1^0$ are the only physical states.  We briefly describe the mean field ansatzes of these three states here, while relegating the details to Appendix \ref{ansatzappendix}. For 2(a) and 2(c), the bond parameters $\chi_{ij}$ and $\Delta_{ij}$ are real, while the onsite term $\lambda_i^{(2)}$ is zero. For $U1^0$, $\chi_{ij}$ is real, while $\Delta_{ij}=0$ by construction.

\subsubsection{$\mathbb{Z}_2$ Spin Liquid State 2(a)}

2(a) is the so-called uniform ansatz. Symmetry-related bond parameters are exactly equal, i.e.~$\chi_{ij}=\chi_{1(2)}$ and $\Delta_{ij}=\Delta_{1(2)}$ for any pair of first (second) nearest neighbors $i$ and $j$, while on-site terms are same at all sites $i$, $\lambda_i^{(3)}=\lambda^{(3)}$ and $\lambda_i^{(1)}=\lambda^{(1)}$. There are in total four independent variational parameters $\lbrace \chi_1, \Delta_1, \chi_2, \Delta_2 \rbrace$, and two Lagrange multipliers $\lbrace \lambda^{(3)}, \lambda^{(1)} \rbrace$ chosen to satisfy the single occupancy constraint.

\subsubsection{$\mathbb{Z}_2$ Spin Liquid State 2(c)}

The mean field ansatz of 2(c) is more elaborate. On-site, first nearest neighbor, and second nearest neighbor hoppings are uniform, i.e.~$\lambda_i^{(3)}=\lambda^{(3)}$ for all sites $i$, and $\chi_{ij}=\chi_{1(2)}$ for all first (second) nearest neighbors $i$ and $j$. On-site and first nearest neighbor pairings are not allowed, i.e.~$\lambda_i^{(1)}=0$ for all sites $i$ and $\Delta_{ij}=0$ for all first nearest neighbors $i$ and $j$. Second nearest neighbor pairings are allowed but admits a sign structure, such that $\Delta_{ij}=\Delta_2$ for some pairs of second nearest neighbors $i$ and $j$, while $\Delta_{ij}=-\Delta_2$ for others. There are in total three independent variational parameters $\lbrace \chi_1, \chi_2, \Delta_2 \rbrace$, and one Lagrange multiplier $\lambda^{(3)}$ chosen to satisfy the single occupancy constraint.

\subsubsection{$U(1)$ Spin Liquid State $U1^0$}
$U1^0$ is the uniform \textit{hopping} ansatz, in which all symmetry-related hopping parameters are exactly equal, i.e.~$\chi_{ij}=\chi_{1(2)}$ for all first (second) nearest neighbors. It can be obtained by turning of the pairing terms in 2(a) or 2(c), i.e. it is the root $U(1)$ state of these $\mathbb{Z}_2$ spin liquids\cite{PhysRevB.95.054404}. For $U(1)$ spin liquids, we do not have to explicitly introduce the onsite terms $\lambda_i^{(3)}$, because the conservation of spinon number implies that, at zero temperature, the single occupancy constraint can be enforced by simply filling the lower half of the energy eigenstates. We will use the terminology ``Fermi level'' $\varepsilon_\mathrm{F}$ in this work, defined as the energy separating the filled and empty states, while noting that it is often identified as the Lagrange multiplier $\lambda^{(3)}$ in the literature. There are in total two independent variational parameters $\chi_1$ and $\chi_2$.

\section{\label{bandsection}Self Consistent Solution and Spinon Spectrum}

Solving the mean field Hamiltonian self consistently, we find that the bond parameters and the Lagrange multipliers of the 2(a) state converge to the form
\begin{subequations}
\begin{align}
(\lambda^{(3)}, \lambda^{(1)}) &= A_0 J_1 (\cos \theta, \sin \theta) , \\
(\chi_1, \Delta_1) &= A_1 (\cos (\theta + \alpha_1), - \sin (\theta + \alpha_1)) , \\
(\chi_2, \Delta_2) &= A_2 (\cos (\theta + \alpha_2), - \sin (\theta + \alpha_2)) .
\end{align}
\end{subequations}
where $A_0=0.0567$, $A_1=0.436$, $A_2=0.330$, $\alpha_1=197^{\circ}$, and $\alpha_2=122^{\circ}$. The solution exhibits a $U(1)$ degree of freedom, $\theta \in [0, 2\pi)$, which arises from the gauge rotation $e^{i \tau_2 \theta/2}$. Such a gauge rotation does not change the previously fixed gauges $g_X$, see Table \ref{hyperkagomepsg}. In addition, the energy remains the same under a sign change of all the hopping (pairing) terms, $\chi_{1,2} \longrightarrow -\chi_{1,2}$ and $\mu_3 \longrightarrow -\mu_3$ ($\Delta_{1,2} \longrightarrow -\Delta_{1,2}$ and $\mu_1 \longrightarrow -\mu_1$), which is generated by the gauge transformation $i \tau_1$ ($i \tau_3$). $i \tau_1$ or $i \tau_3$ only alters the previously fixed gauges $g_X$ by at most a minus sign, which is inconsequential as the IGG is $\mathbb{Z}_2$\cite{PhysRevB.95.054404}.

For the 2(c) state, the convergent solution is
\begin{equation} \label{2csolution}
\begin{aligned}[b]
& \lambda^{(3)}=-0.0617 J_1, \, \chi_1=-0.0504, \\
& \chi_2=0.435, \, \Delta_2=0.0538 .
\end{aligned}
\end{equation}
Gauge equivalent solutions are generated by the transformations $i \tau_1$, $i \tau_2$, and $i \tau_3$, which change the previously fixed gauges $g_X$ by at most a minus sign while preserving the energy. $i \tau_1$ ($i \tau_3$) flips the signs of all the hopping (pairing) terms. $i \tau_2$ flips the sign of all the hopping \textit{and} pairing terms.

For the $U1^0$ state, the convergent solution is
\begin{equation} \label{u10solution}
\varepsilon_\mathrm{F}=0.0574 J_1, \, \chi_1=-0.0493, \chi_2=0.438.
\end{equation}
where $\varepsilon_\mathrm{F}$ is the Fermi level determined by half filling. Gauge equivalent solutions are generated by transformations of the form $i \tau_1 e^{i \theta \tau_3}$, where $\theta \in [0, 2\pi)$. These transformations change $g_\mathcal{T}$ to $g_\mathcal{T} e^{2 i \theta \tau_3}$, which is inconsequential as the IGG is $U(1)$, while leaving all other previously fixed gauges invariant. $i \tau_1 e^{i \theta \tau_3}$ changes the signs of all terms in \eqref{u10solution}, which can be viewed as a particle-hole transformation.

\begin{figure}
\includegraphics[scale=0.25]{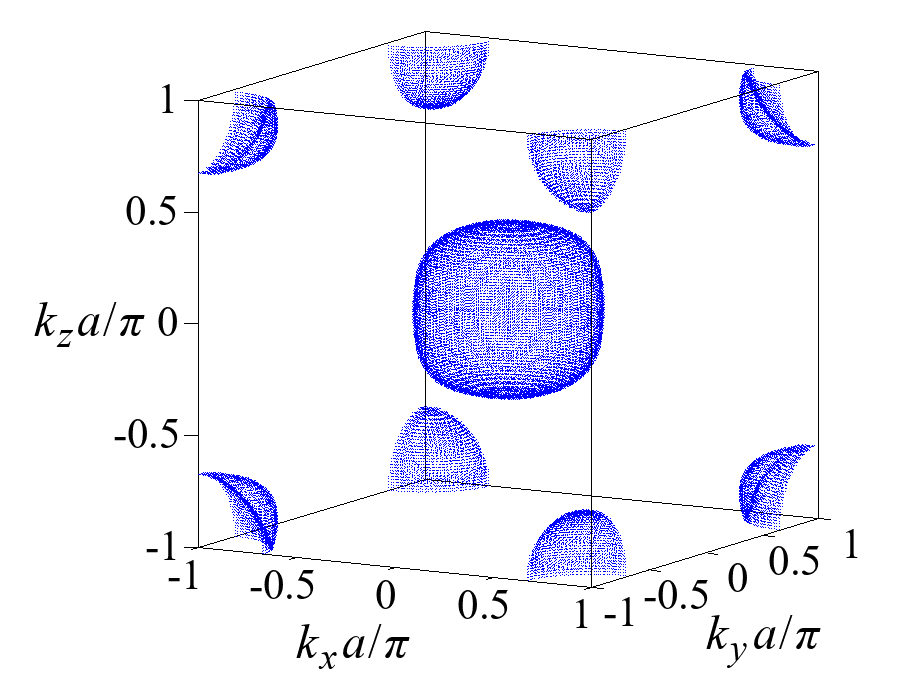}
\caption{\label{fermisurface}The $U(1)$ spin liquid $U1^0$ exhibits a Fermi surface of spinons.}
\end{figure}

Upon convergence, we find that, up to three significant figures, the energies per site of 2(a), 2(c), and $U1^0$ are $-0.102 J_1$, $-0.0966 J_1$, and $-0.0966 J_1$ respectively. Therefore, 2(a) is the ground state at the mean field level. 2(c) and $U1^0$ are very close in energy\cite{energydifference} as their solutions are similar - they are characterized by a dominant second nearest neighbor hopping $\chi_2$, with other terms at the subleading order, see $\eqref{2csolution}$ and $\eqref{u10solution}$. We further examine the spinon spectra of these states. $U1^0$ is gapless with a spinon Fermi surface, as shown in Fig.~\ref{fermisurface}. 2(a) is gapped with a small excitation gap of $0.0259 J_1$. 2(c) is proximate to $U1^0$, but some portions of the spinon Fermi surface may be gapped out by the finite pairing term $\Delta_2$. Since $\Delta_2$ is small and we find energies down to the scale $10^{-6} J_1$ from numerics, we can say that 2(c) is practically gapless. The collection of low energy excitations in 2(c) resembles the spinon Fermi surface in $U1^0$.

\begin{figure}
\subfloat[]{\label{2aJ1J2dispersion}
\includegraphics[scale=0.14]{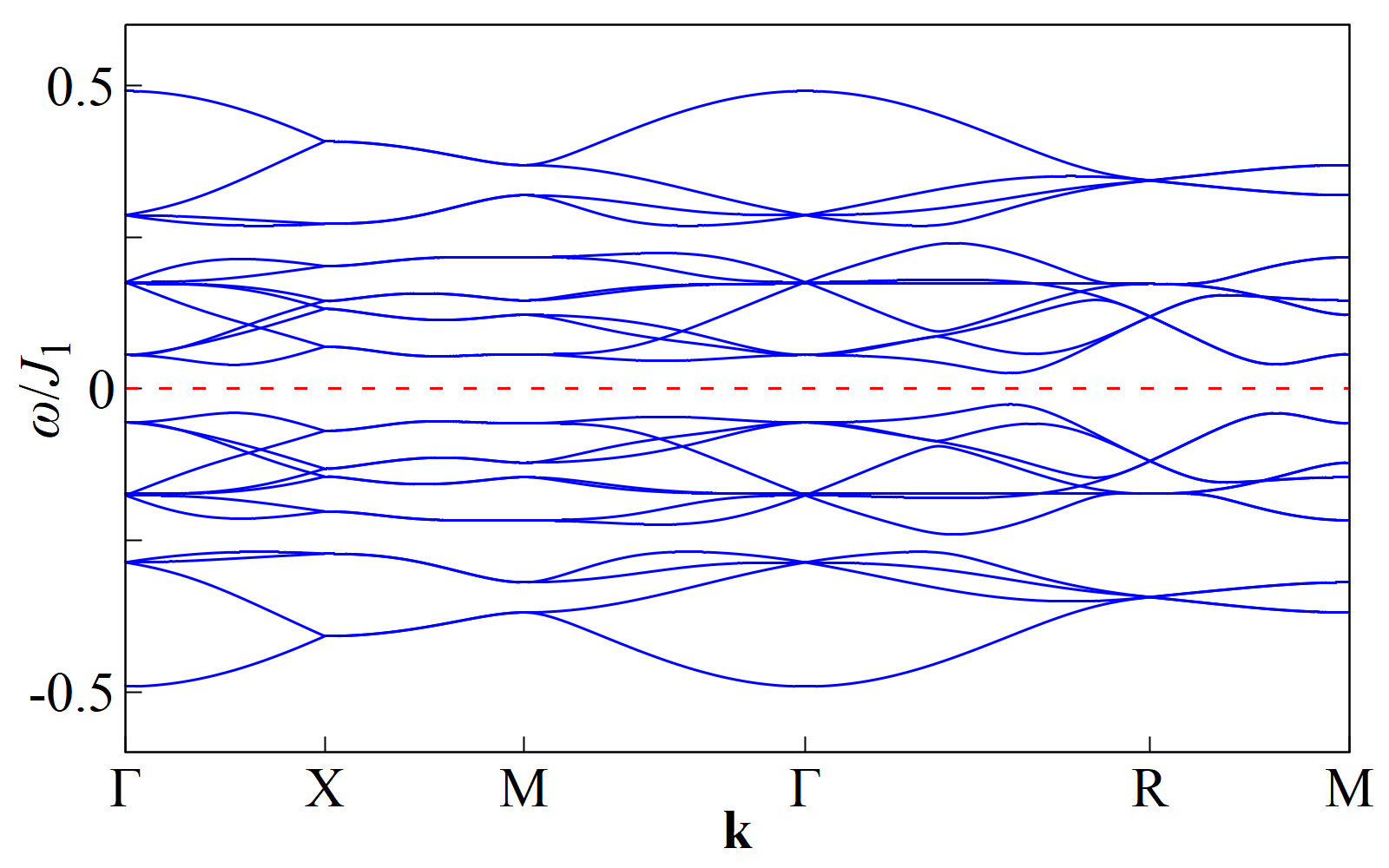}} \\
\subfloat[]{\label{2cJ1J2dispersion}
\includegraphics[scale=0.14]{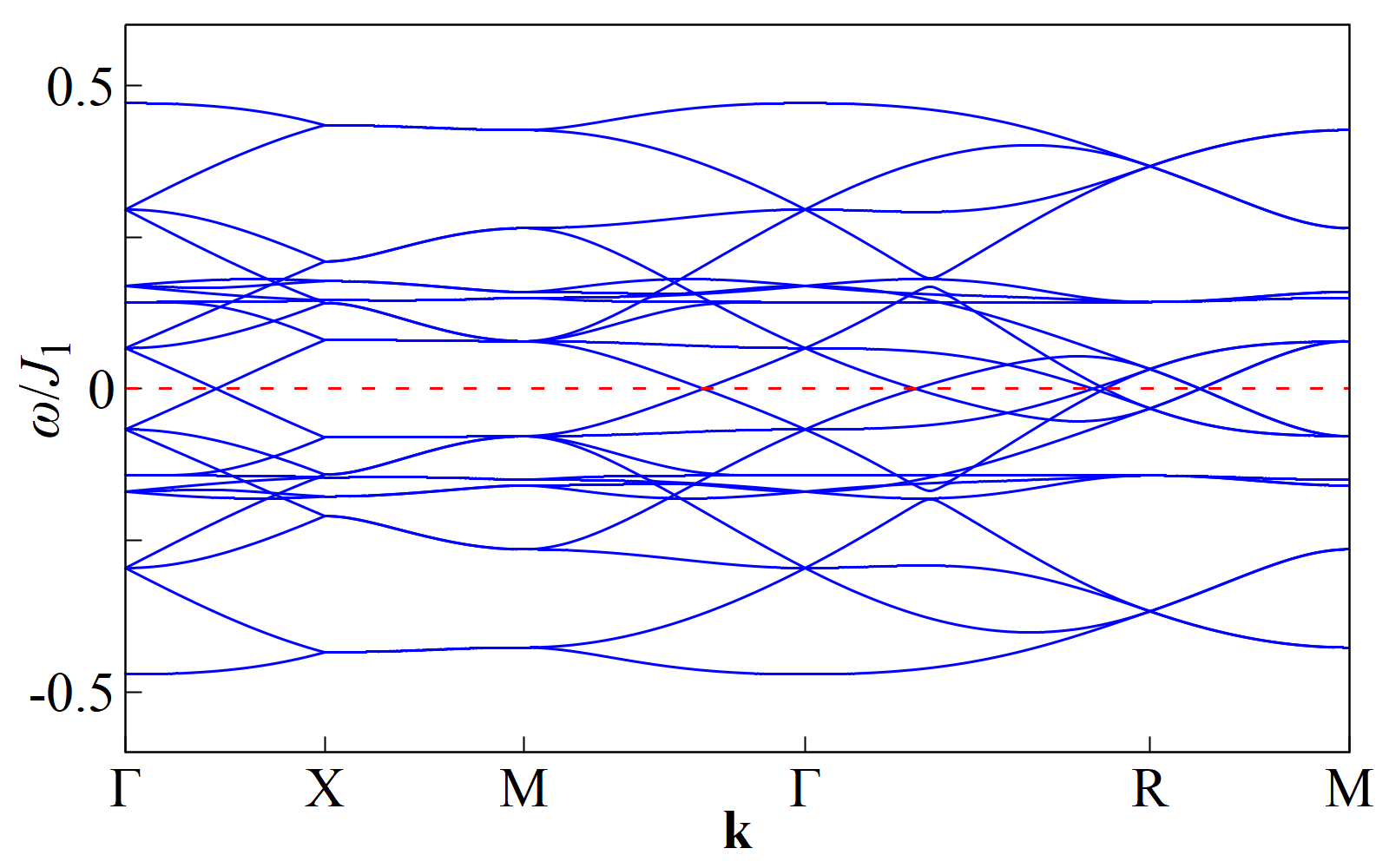}} \\
\subfloat[]{\label{u10J1J2dispersion}
\includegraphics[scale=0.14]{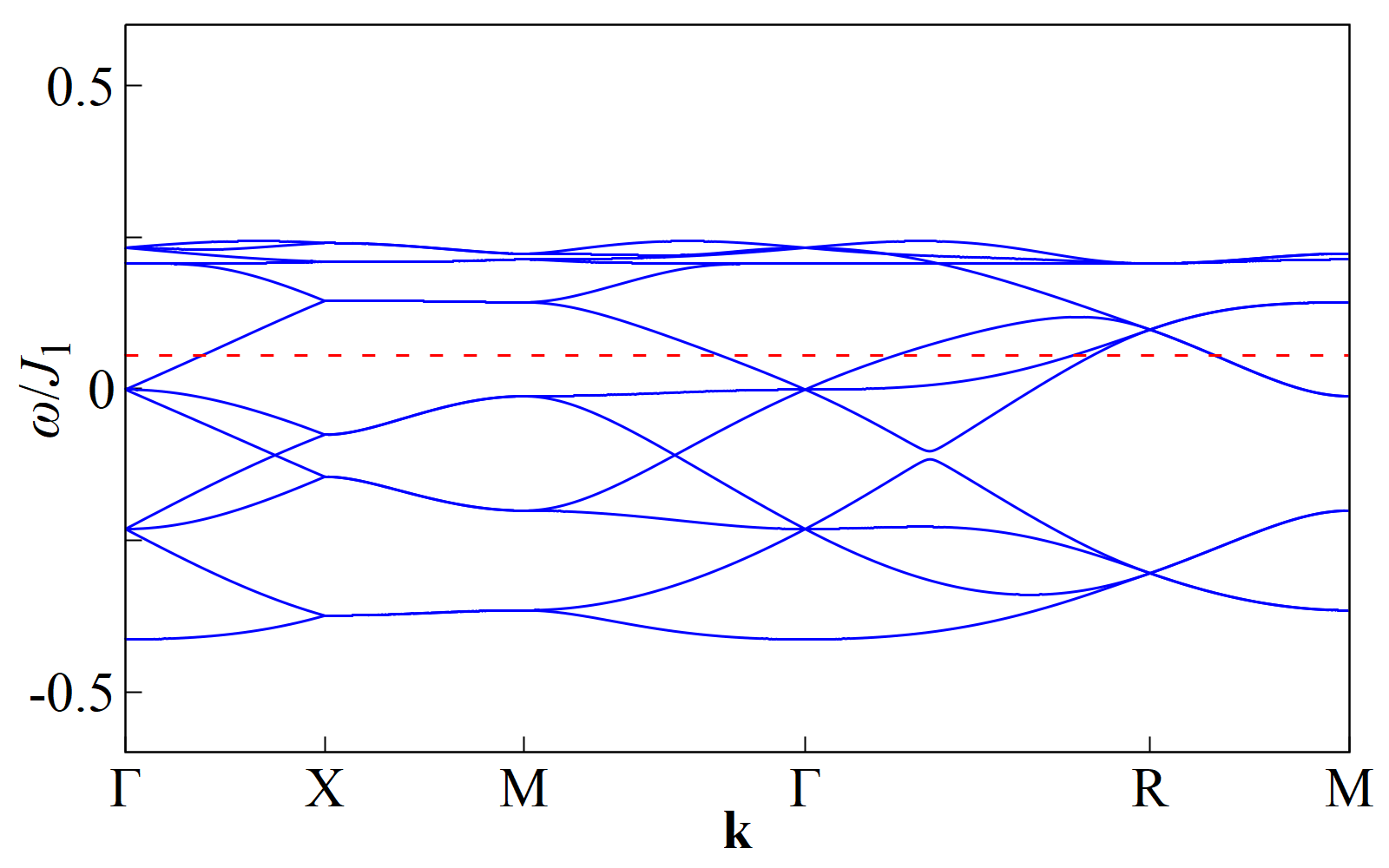}}
\caption{Spinon Dispersions of the $\mathbb{Z}_2$ spin liquid states (a) 2(a) and (b) 2(c), and (c) the $U(1)$ spin liquid state $U1^0$. The high symmetry momenta are $\Gamma=(0,0,0)$, $\mathrm{X}=(1,0,0)\pi/a$, $\mathrm{M}=(1,1,0)\pi/a$, and $\mathrm{R}=(1,1,1)\pi/a$. In (a) and (b), the red dashed lines indicate the zero energy. In (c), the red dashed line indicates the Fermi level, below (above) which the states are filled (empty), and each band is two-fold degenerate.}
\end{figure}

The spinon dispersions of 2(a), 2(c), and $U1^0$ are plotted along high symmetry directions in the first Brillouin zone\cite{SETYAWAN2010299}, see Figs.~\ref{2aJ1J2dispersion}, \ref{2cJ1J2dispersion}, and \ref{u10J1J2dispersion} respectively.

\section{\label{heatsection}Heat Capacity}

\begin{figure}
\subfloat[]{\label{2aheat}
\includegraphics[scale=0.16]{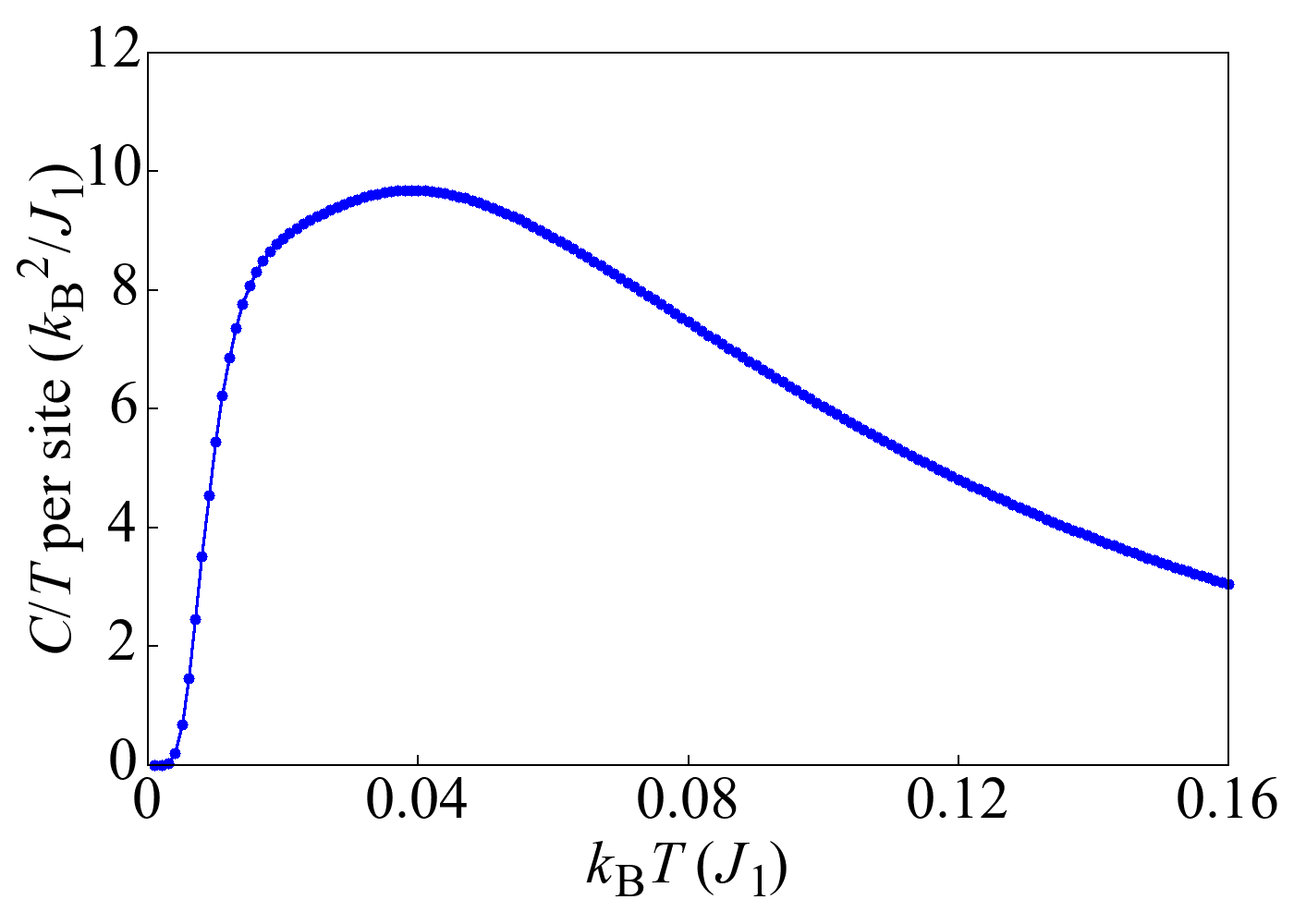}} \\
\subfloat[]{\label{2cheat}
\includegraphics[scale=0.16]{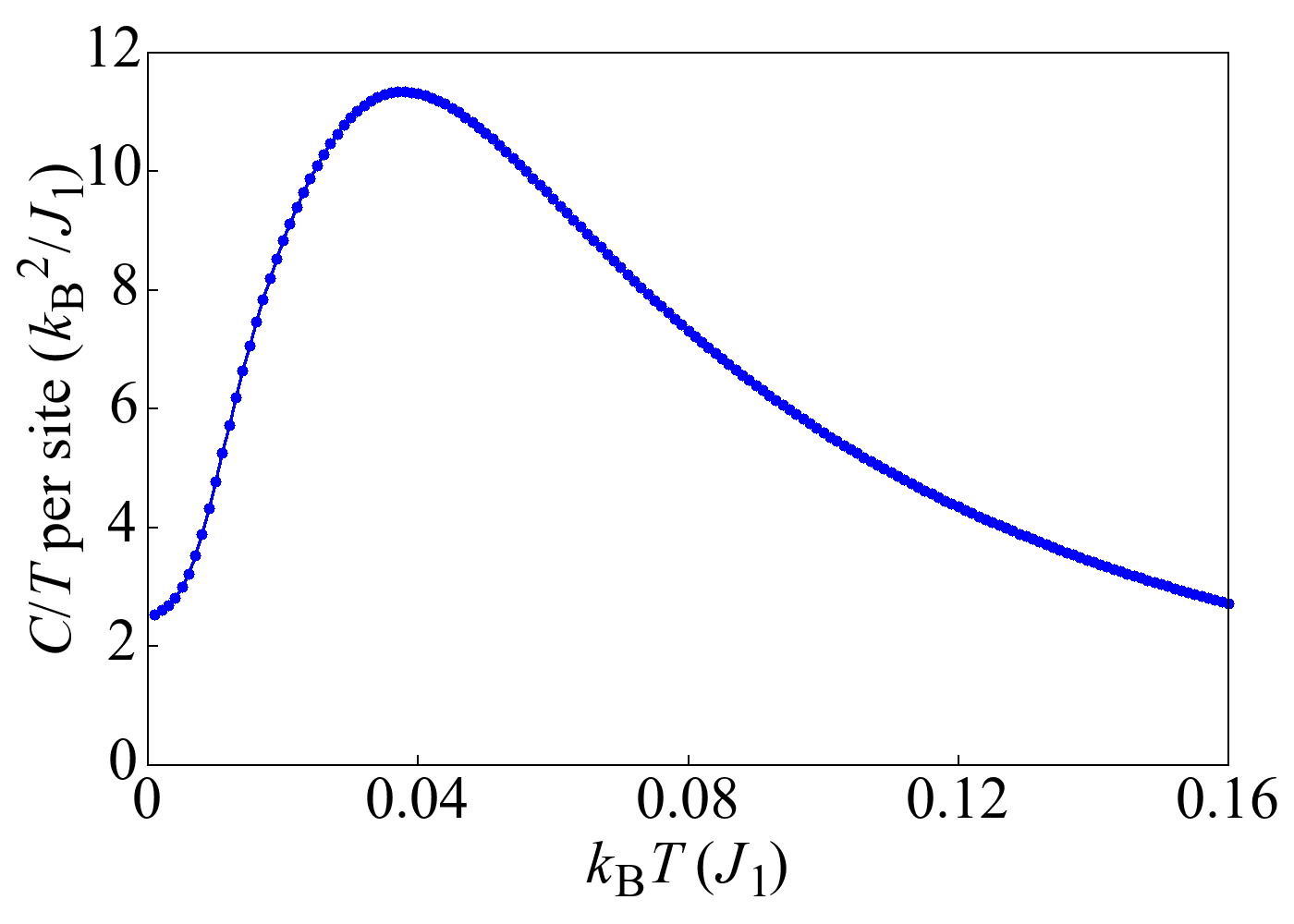}}
\caption{Heat capacity coefficients $C/T$ per site as a function of temperature $T$ of the two $\mathbb{Z}_2$ spin liquid states (a) 2(a) and (b) 2(c).}
\end{figure}

Heat capacity should be able to distinguish phases with gapped and gapless quasiparticle excitations. We illustrate this for the $\mathbb{Z}_2$ spin liquids 2(a) and 2(c). As discussed in Appendix \ref{hamiltonianappendix}, we interpret the Bogoliubov quasiparticles as carrying non-negative energies $\omega_{\mathbf{k} \uparrow}, \omega_{-\mathbf{k} \downarrow} \geq 0$, where the sublattice index has been suppressed for brevity. Furthermore, since $\omega_{\mathbf{k} \uparrow} = \omega_{-\mathbf{k} \downarrow}$, we can drop the spin index and write both of them as $\omega_\mathbf{k}$. Heat capacity is given by the derivative of the total energy with respect to temperature. Using \eqref{diagonalhamiltonian},
\begin{equation}
\begin{aligned}[b]
C &= \frac{\partial \langle H \rangle}{\partial T} = 2 \sum_\mathbf{k} \frac{\partial}{\partial T} \omega_\mathbf{k} n (\omega_\mathbf{k}) \\
&= \frac{2}{k_\mathrm{B} T^2} \sum_\mathbf{k} \omega_\mathbf{k}^2 \frac{e^{\omega_\mathbf{k}/k_\mathrm{B}T}}{(e^{\omega_\mathbf{k}/k_\mathrm{B}T}+1)^2} ,
\end{aligned}
\end{equation}
where $n(\omega)=[1+\exp(\omega/k_\mathrm{B}T)]^{-1}$ is the Fermi-Dirac distribution, and we have also assumed that the temperature scale is low enough such that the spinon spectrum remains the same. We plot the heat capacity coefficients $C/T$ per site for 2(a) and 2(c) in Figs.~\ref{2aheat} and \ref{2cheat} respectively. In the zero temperature limit, $C/T$ of 2(a) vanishes, while that of 2(c) is finite. However, real experiments may not be able to access this very low temperature regime, given the interaction energy scale of $J_1 \approx 1 \, \mathrm{meV}$. At higher temperatures, the distinction between the two is not so obvious. They both show a broad peak similar to what is observed in the experiment\cite{PhysRevB.90.035141}.

\section{\label{dynamicsection}Dynamical Spin Structure Factor}

Ref.~\onlinecite{s41467-020-15594-1} reports the inelastic neutron scattering (INS) spectra in the $[h,h,l]$ and $[h,k,0]$ planes at the energy $E=0.5 \, \mathrm{meV}$ (integrated over certain ranges of momenta and energies), see Figs.~4a and 4b in Ref.~\onlinecite{s41467-020-15594-1}. Several INS data in the $[h,k,0]$ plane at higher energies are also provided, see Figs.~3e-3g in Ref.~\onlinecite{s41467-020-15594-1}. Notably, a diffusive ring-like structure is observed in the $[h,k,0]$ plane, with an approximate radius of $\lvert \mathbf{k} \rvert \approx 0.8 / \textrm{\AA} \approx 1.6 \, \mathrm{r.l.u.}$ The ring is most obvious at lower energies $E=0.5$ and $0.75 \, \mathrm{meV}$, but gradually weakens and merges into the background at higher energies $E=1.5$ and $2 \, \mathrm{meV}$. We remark that the ring is not perfectly isotropic (i.e.~having the same intensity in every radial direction), but exhibits a four-fold rotational symmetry as one would expect from a cubic space group.

\begin{figure*}
\subfloat[]{\label{2adynamichhlw060d002}
\includegraphics[scale=0.16]{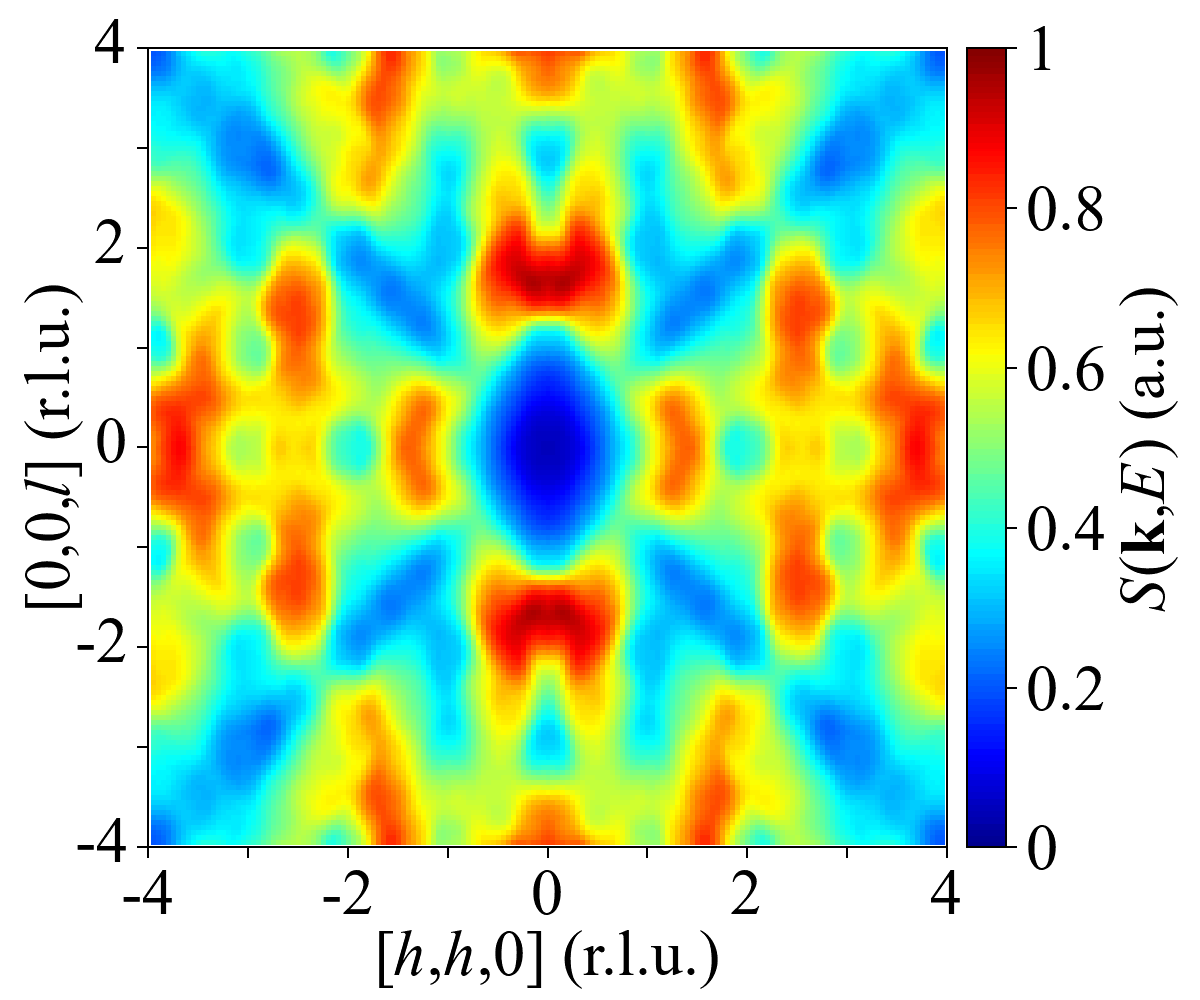}} \quad
\subfloat[]{\label{2cdynamichhlw060d002}
\includegraphics[scale=0.16]{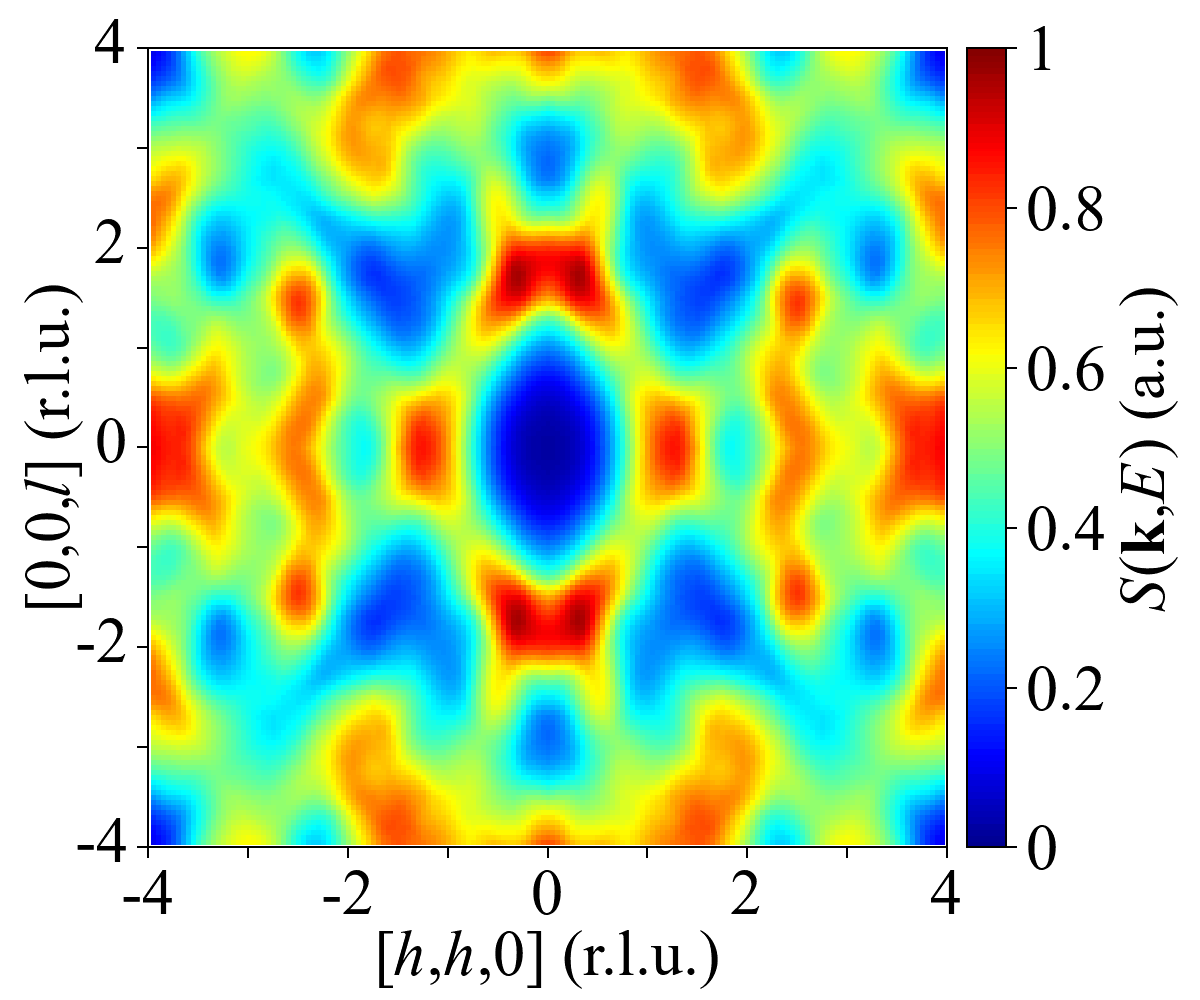}} \quad
\subfloat[]{\label{u10dynamichhlw060d002}
\includegraphics[scale=0.16]{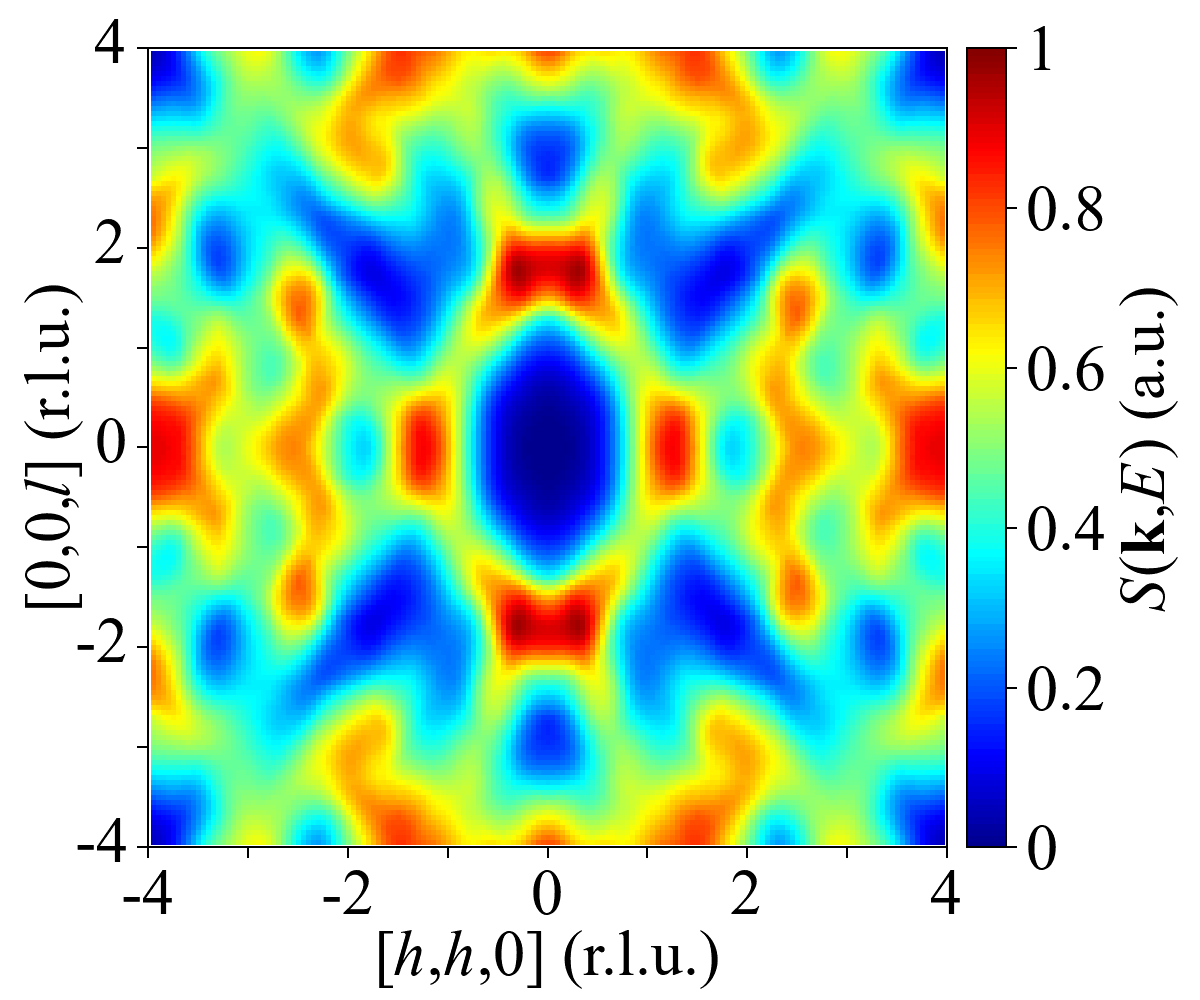}} \\
\subfloat[]{\label{2adynamichk0w060d002}
\includegraphics[scale=0.16]{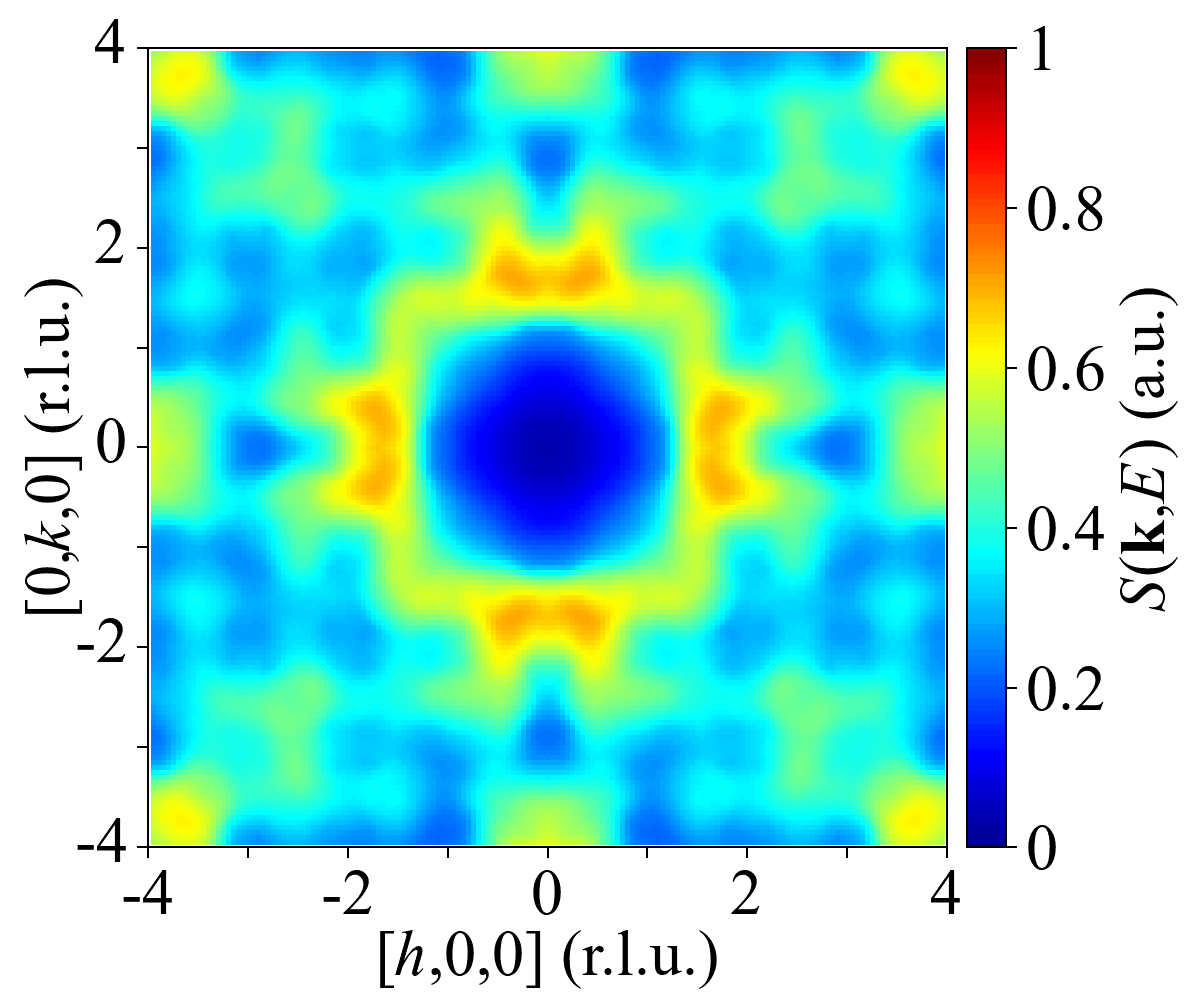}} \quad
\subfloat[]{\label{2cdynamichk0w060d002}
\includegraphics[scale=0.16]{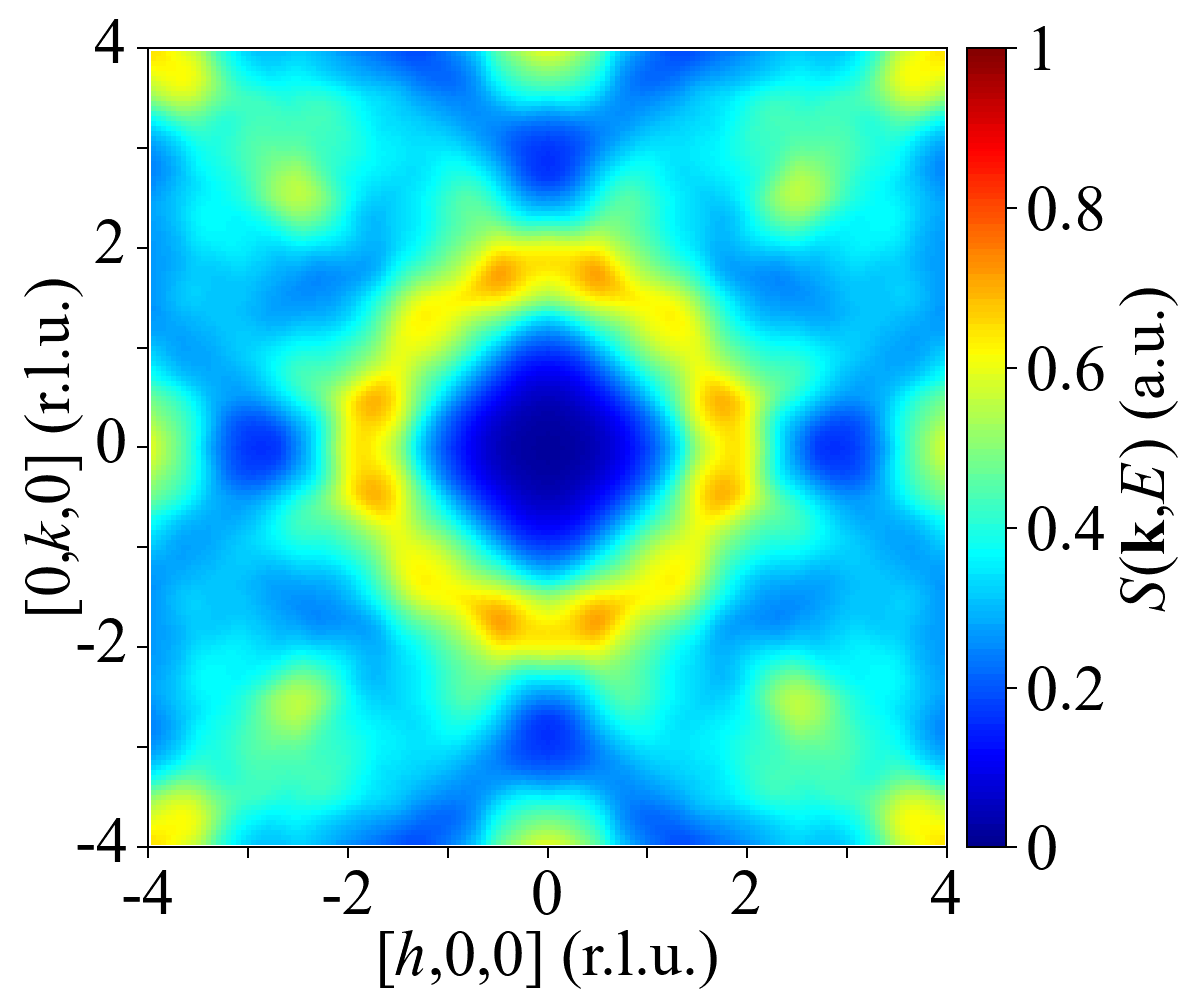}} \quad
\subfloat[]{\label{u10dynamichk0w060d002}
\includegraphics[scale=0.16]{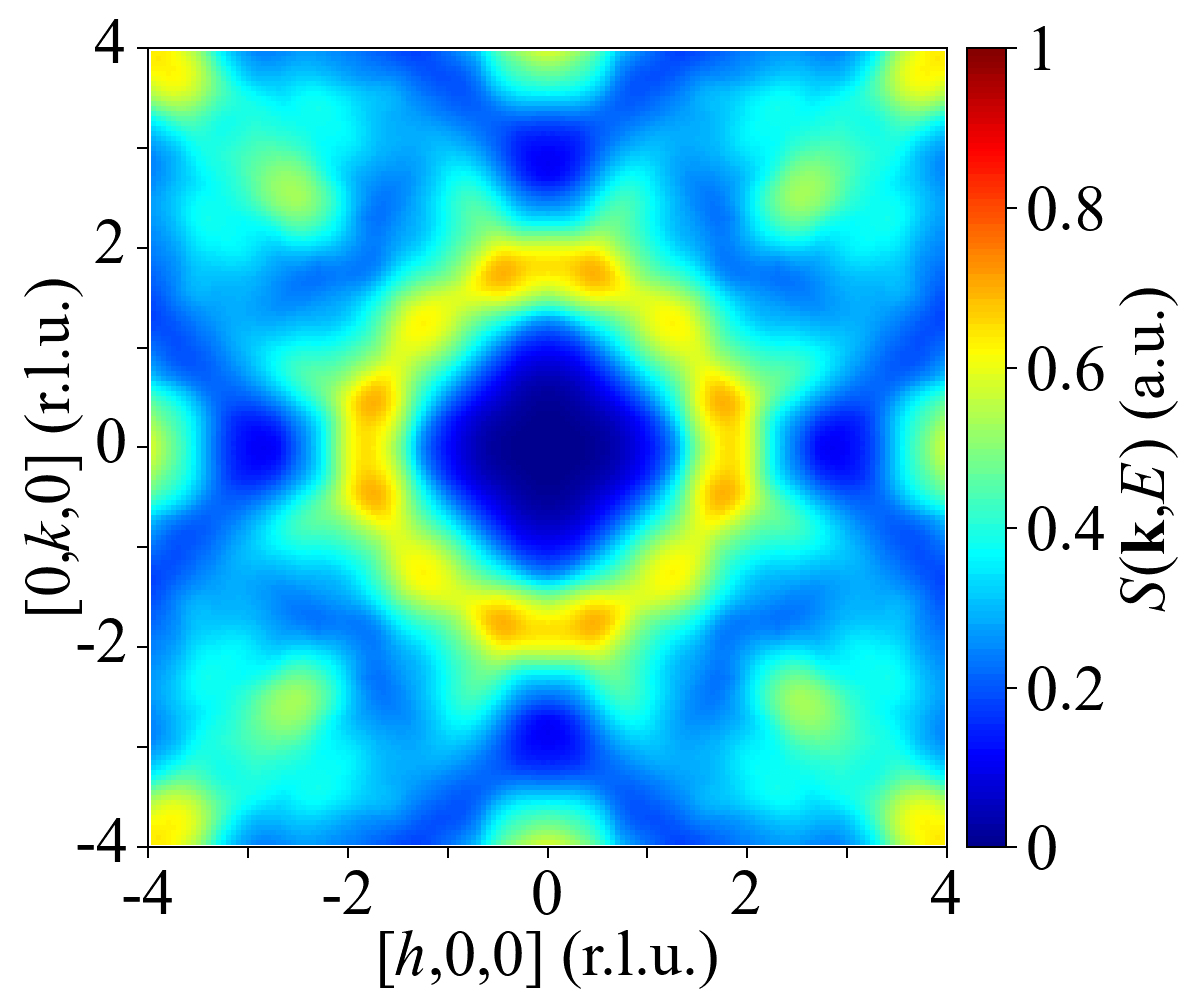}}
\caption{Dynamical spin structure factors of the three candidate spin liquid states, plotted in the $[h,h,l]$ and $[h,k,0]$ planes (upper and lower panels), at $E=0.6 J_1$ with $\Delta E=0.02 J_1$. (a,d) The $\mathbb{Z}_2$ spin liquid 2(a). (b,e) The $\mathbb{Z}_2$ spin liquid 2(c). (c,f) The $U(1)$ spin liquid $U1^0$. These subplots can be compared to Figs.~4a and 4b in Ref.~\onlinecite{s41467-020-15594-1}. As in the experiment, the intensities are given in arbitrary units (a.u.); the color scale $[0,1]$ in each subplot is relative but not absolute. The momenta are measured in reciprocal lattice unit, $1 \, \mathrm{r.l.u.}= 2 \pi /a$ where $a$ is the lattice constant.}
\end{figure*}

\begin{figure*}
\subfloat[]{\label{2adynamichhlw070d010}
\includegraphics[scale=0.16]{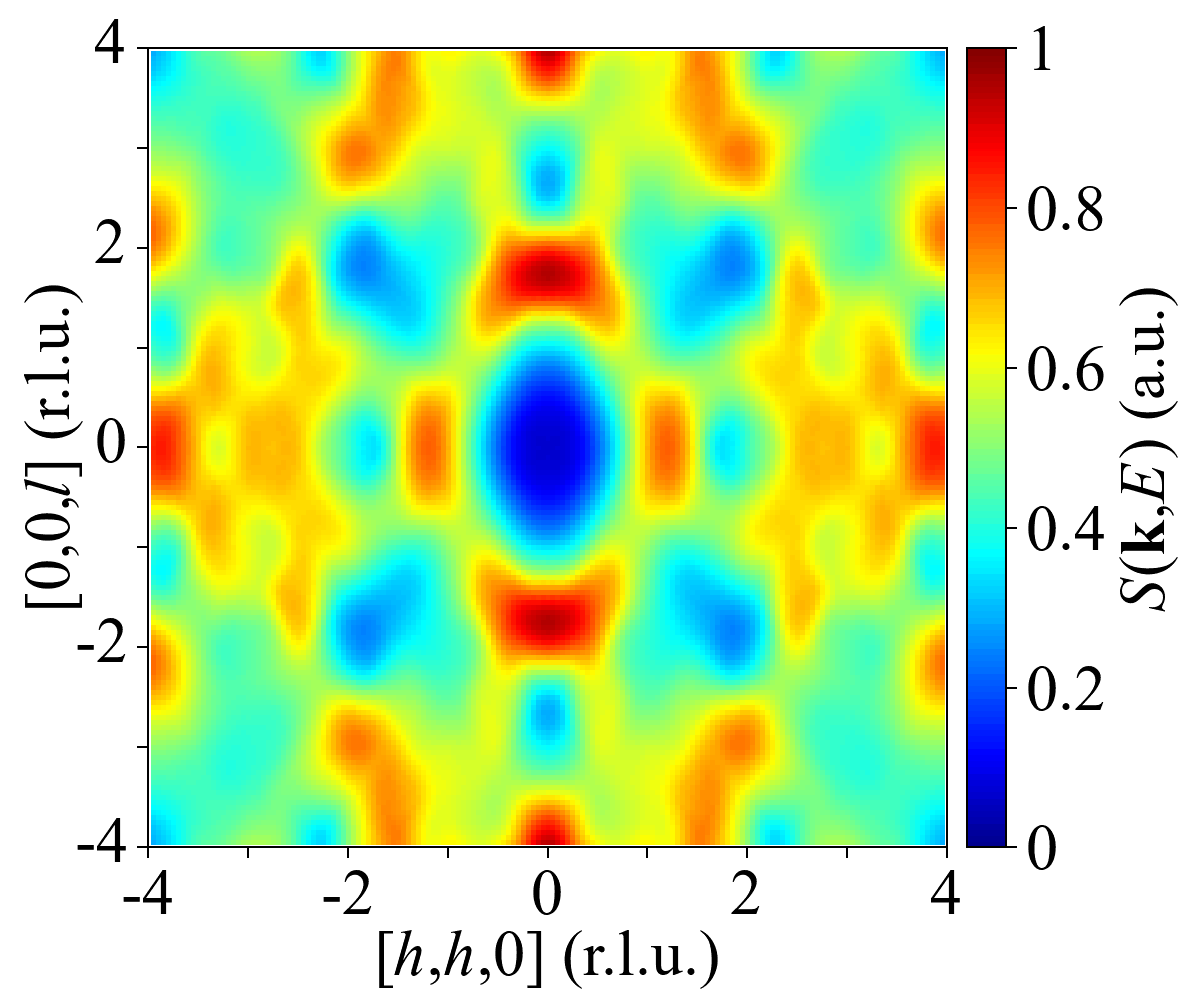}} \quad
\subfloat[]{\label{2cdynamichhlw070d010}
\includegraphics[scale=0.16]{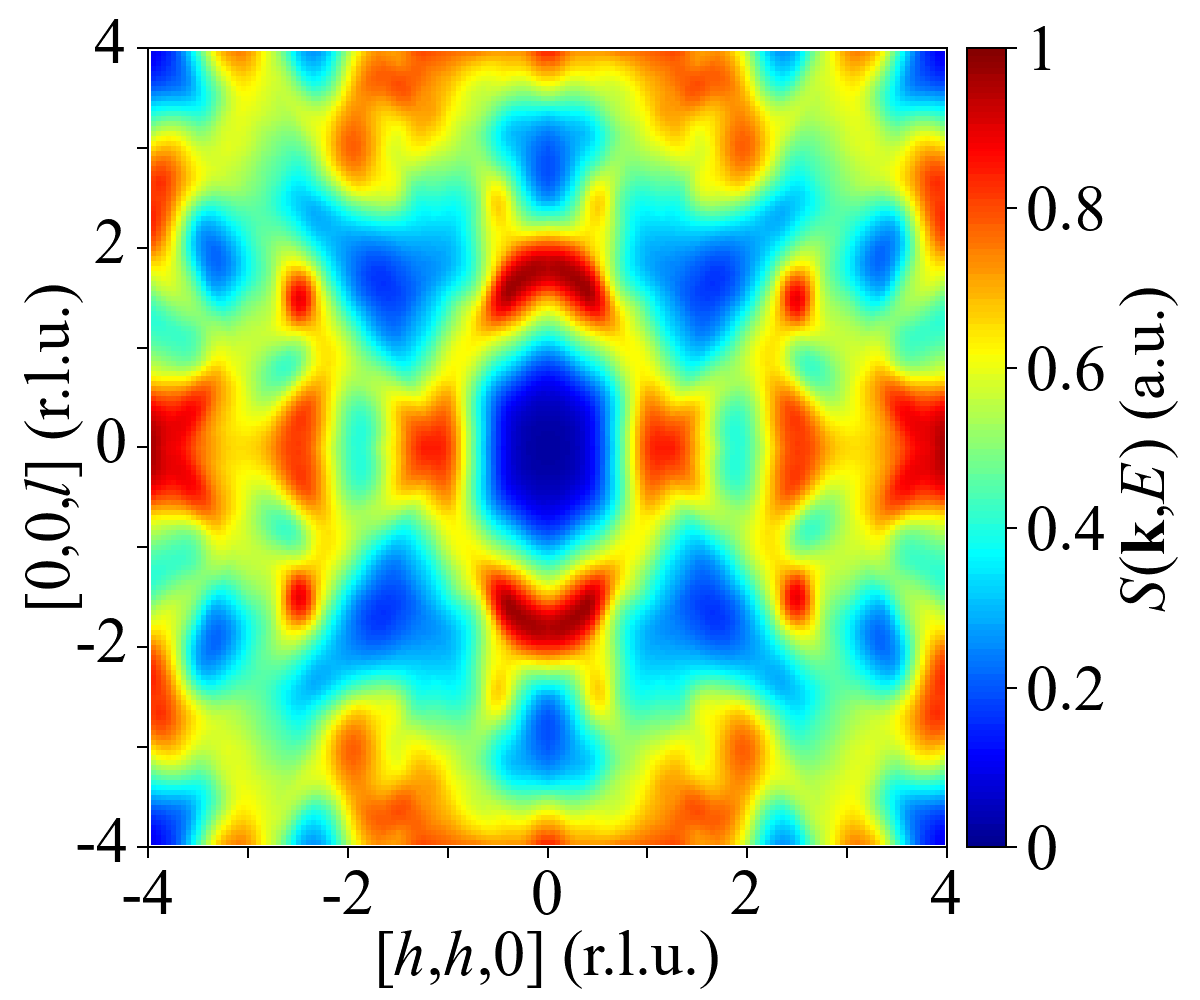}} \quad
\subfloat[]{\label{u10dynamichhlw070d010}
\includegraphics[scale=0.16]{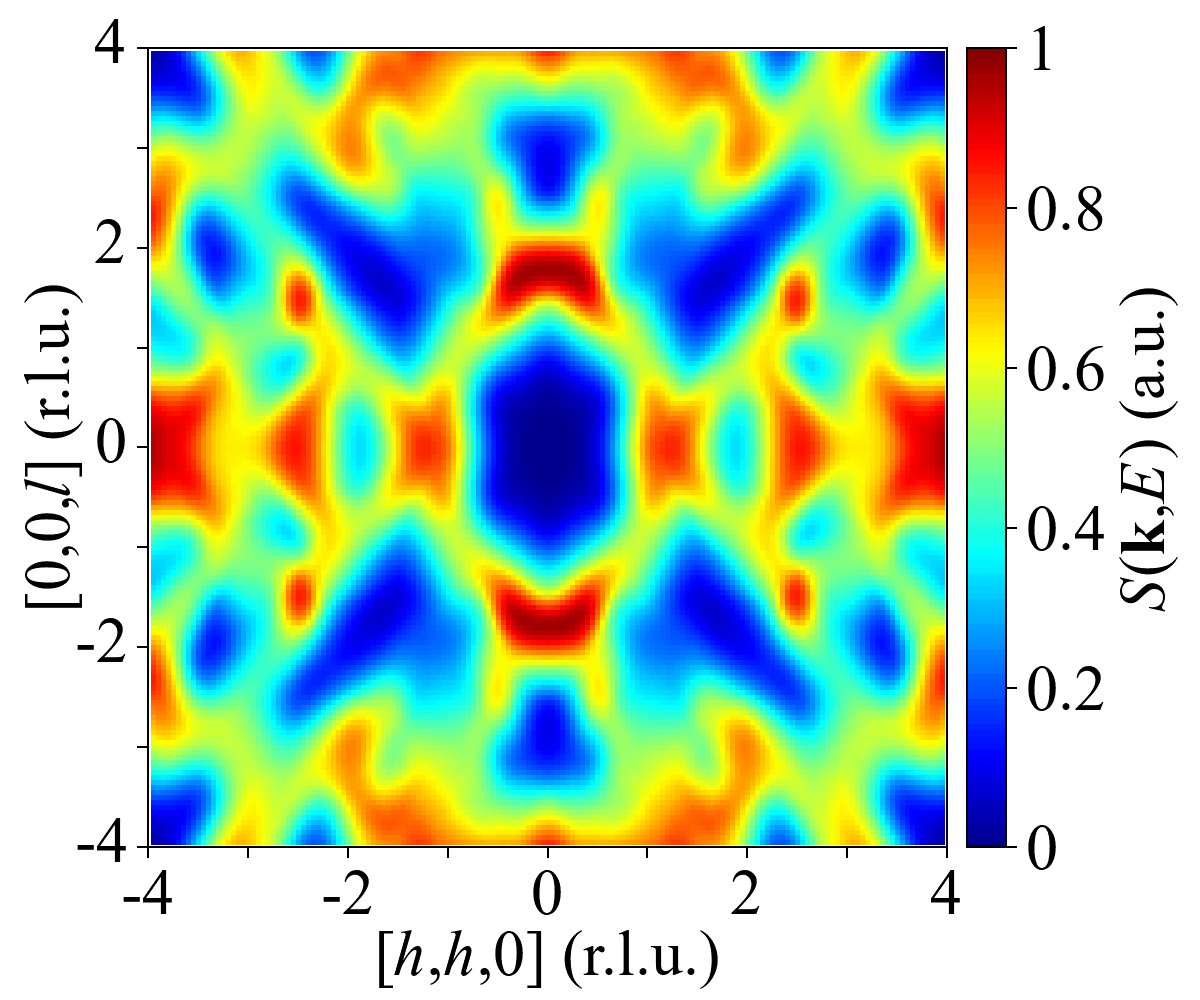}} \\
\subfloat[]{\label{2adynamichk0w070d010}
\includegraphics[scale=0.16]{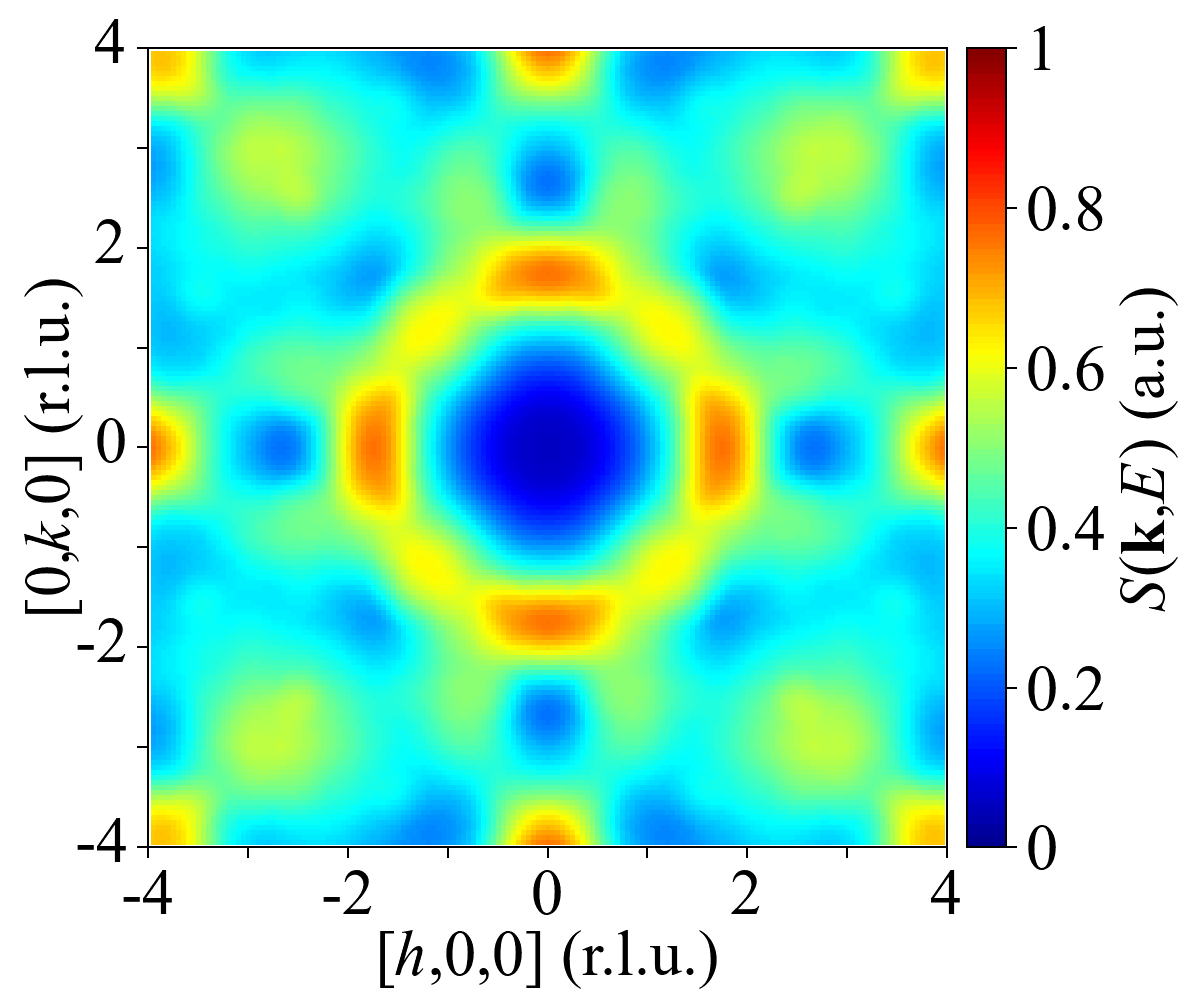}} \quad
\subfloat[]{\label{2cdynamichk0w070d010}
\includegraphics[scale=0.16]{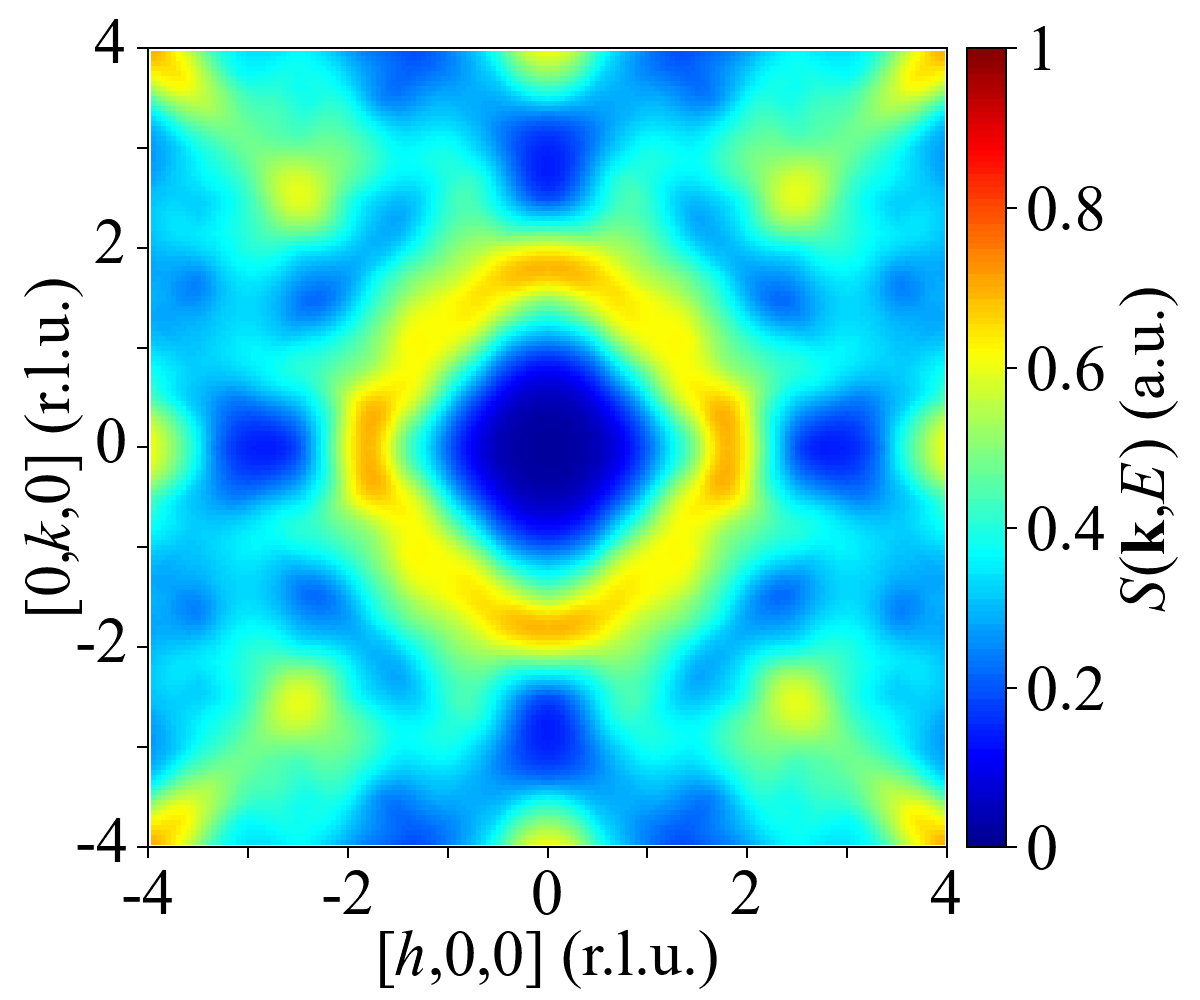}} \quad
\subfloat[]{\label{u10dynamichk0w070d010}
\includegraphics[scale=0.16]{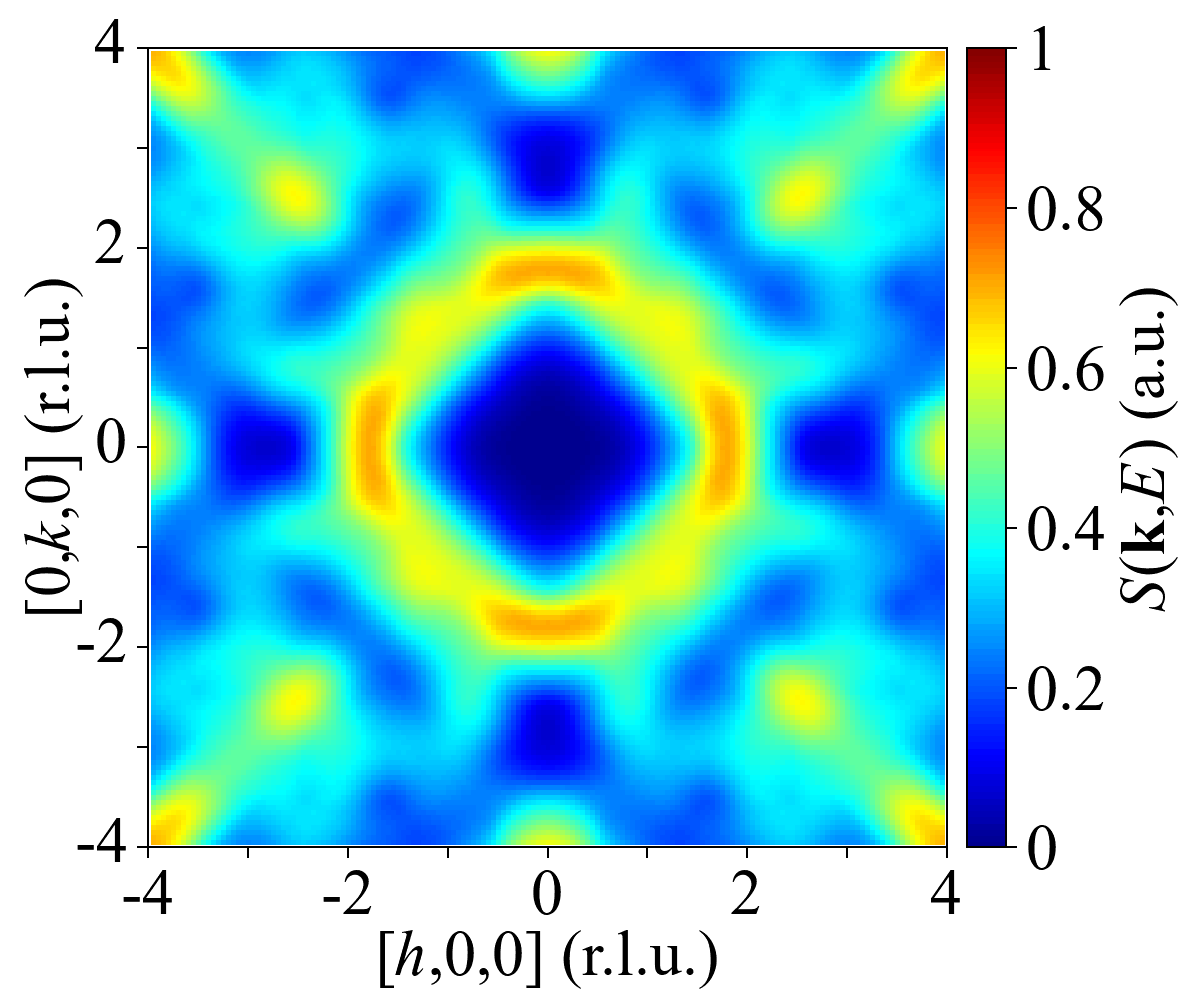}}
\caption{Dynamical spin structure factors of the three candidate spin liquid states, plotted in the $[h,h,l]$ and $[h,k,0]$ planes (upper and lower panels), at $E=0.7 J_1$ with $\Delta E=0.1 J_1$. (a,d) The $\mathbb{Z}_2$ spin liquid 2(a). (b,e) The $\mathbb{Z}_2$ spin liquid 2(c). (c,f) The $U(1)$ spin liquid $U1^0$. These subplots can be compared to Figs.~4a and 4b in Ref.~\onlinecite{s41467-020-15594-1}. As in the experiment, the intensities are given in arbitrary units (a.u.); the color scale $[0,1]$ in each subplot is relative but not absolute. The momenta are measured in reciprocal lattice unit, $1 \, \mathrm{r.l.u.}= 2 \pi /a$ where $a$ is the lattice constant.}
\end{figure*}

To compare with the INS experiments, we calculate the dynamical spin structure factor
\begin{equation} \label{dynamic}
\mathcal{S} \left( \mathbf{k}, E \right) = \sum_{ij} \int \mathrm{d}t \, e^{i E t} \left \langle \mathbf{S}_i (t) \cdot \mathbf{S}_j (0) \right \rangle e^{i \mathbf{k} \cdot (\mathbf{r}_j-\mathbf{r}_i)}
\end{equation}
for the three candidate spin liquids in the $[h,h,l]$ and $[h,k,0]$ planes. Details of the calculation are provided in Appendix \ref{dynamicappendix}. We examine slices of $\mathcal{S} (\mathbf{k}, E)$ at several energies $E$. For each $E$, we integrate $\mathcal{S}(\mathbf{k},E)$ over a small window $(E - \Delta E, E + \Delta E)$ of energies centered at $E$. What we mean by $\mathcal{S} (\mathbf{k},E)$ in the following is actually $\int_{E-\Delta E}^{E+\Delta E} \mathrm{d} E' \, \mathcal{S} (\mathbf{k}, E')$.

We find that the dynamical spin structure factors of the 2(c) state at $E = 0.6 J_1$ highly resembles the experimentally observed INS spectra at $E = 0.5 \, \mathrm{meV}$ (Figs.~4a and 4b in Ref.~\onlinecite{s41467-020-15594-1}). One can see that the intensity distribution in the $[h,h,l]$ plane (Fig.~\ref{2cdynamichhlw060d002}) is similar to the INS data, especially considering the shapes and locations of the strongest signals. In the $[h,k,0]$ plane, we also observe a diffusive ring like structure (Fig.~\ref{2cdynamichk0w060d002}) with an approximate radius of $\lvert \mathbf{k} \rvert=1.8 \, \mathrm{r.l.u.}$ and pairs of maxima, which can be compared to the experiment. It is not surprising that the dynamical spin structure factors of the $U1^0$ state (Figs.~\ref{u10dynamichhlw060d002} and \ref{u10dynamichk0w060d002}) look very much like those of 2(c), and thus the INS spectra, as these states are proximate to each other. For the 2(a) state, $\mathcal{S} (\mathbf{k},E)$ at $E=0.6 J_1$ is similar to the INS spectrum in the $[h,h,l]$ plane (Fig.~\ref{2adynamichhlw060d002}), but not quite in the $[h,k,0]$ plane (Fig.~\ref{2adynamichk0w060d002}), where the high intensity region looks more like a square than a ring. However, the similarity can be improved by going to other energies $E$ and/or changing the width of the integral $\Delta E$, see for example Figs.~\ref{2adynamichhlw070d010} and \ref{2adynamichk0w070d010}.

Density functional theory estimates the interaction energy scale to be $J_1 \approx 1 \, \mathrm{meV}$\cite{s41467-020-15594-1}. Therefore, the energy $E=0.6 J_1$ at which we calculate the dynamical spin structure factor is approximately $0.6 \, \mathrm{meV}$. However, since our calculation is based on a mean field approximation, we do not expect that the energy scale of our theory can be directly compared to that of the experiment. We simply remark that the given INS spectra at $E=0.5 \, \mathrm{meV}$ is similar to $\mathcal{S} (\mathbf{k},E)$ calculated in the range $0.6 J_1 \lesssim E \lesssim 0.7 J_1$, which, evaluated at $J_1 \approx 1 \, \mathrm{meV}$, are not very far from $0.5 \, \mathrm{meV}$.

\section{\label{discussion}Discussion}

In this work, we consider the $J_1$-$J_2$ model of PbCuTe$_2$O$_6$ with antiferromagnetic $J_1 = J_2$, which results in a three dimensional structure of corner sharing triangles known as the hyper-hyperkagome lattice\cite{s41467-020-15594-1}. It exhibits a richer connectivity than the hyperkagome lattice\cite{PhysRevLett.99.137207,PhysRevB.95.054404}, but they belong to the same space group P$4_132$. This allows us to extend the established PSG analysis of the latter\cite{PhysRevB.95.054404} to the former, within the framework of the complex fermion mean field theory\cite{PhysRevB.44.2664,PhysRevB.49.5200,PhysRevB.65.165113,PhysRevB.83.224413}. We demonstrate that only two out of the five possible $\mathbb{Z}_2$ spin liquids, and one out of the two possible $U(1)$ spin liquids, yield physical ansatzes on the hyper-hyperkagome structure. They are labelled as 2(a), 2(c), and $U1^0$ respectively, with 2(a) being the lower energy state at the mean field level. 2(a) is gapped, while 2(c) and $U1^0$ are gapless and proximate to each other. The calculated dynamical spin structure factors of these spin liquid states are very similar to the INS spectra of PbCuTe$_2$O$_6$\cite{s41467-020-15594-1}. We further show that the gapped and gapless spin liquids can, in principle, be distinguished by heat capacity at very low temperatures, using 2(a) and 2(c) as examples.

Our $J_1$-$J_2$ model is a simplified version of the more elaborate $J_1$-$J_2$-$J_3$-$J_4$ model proposed by the most recent DFT calculation, which uses an energy mapping approach\cite{s41467-020-15594-1}. The pseudo-fermion functional renormalization group (PFFRG) analysis has been applied to the $J_1$-$J_2$-$J_3$-$J_4$ model in Ref.~\onlinecite{s41467-020-15594-1}, which does not find any long range magnetic order at the lowest temperatures. The static spin susceptilibity calculated by PFFRG also highly resembles the INS data. However, PFFRG can only suggests a spin liquid ground state from the absence of magnetic order. In this work, we explicitly identify three candidate spin liquid states which result from the PSG analysis, and show that the corresponding dynamical spin structure factors are in good agreement with the INS data. Our results indicate that one of these spin liquid states may be realized in PbCuTe$_2$O$_6$.

An earlier DFT calculation, which uses a perturbation theory approach, proposes a $J_1$-$J_2$-$J_3$ model for PbCuTe$_2$O$_6$ with $J_1:J_2:J_3=0.54:1:0.77$\cite{PhysRevB.90.035141}. In this model, the $J_2$ interaction, which connects the sites into a hyperkagome network, is dominant. Classical Monte Carlo simulations and Schwinger boson mean field theory have been applied to this $J_1$-$J_2$-$J_3$ model, with varying strengths of $J_1$ and $J_3$, to investigate the possible magnetic orders and quantum spin liquids in Ref.~\onlinecite{PhysRevB.101.054408}. A PSG analysis similar to Ref.~\onlinecite{PhysRevB.95.054404} has been carried out to classify the bosonic spin liquids. While the Schwinger boson approach enables one to interpolate between the classical and quantum limits, it excludes the case of a stable gapless spin liquid, as the condensation of spinons leads to a magnetically ordered state. Given the experimental indications of a spin liquid that is gapless or has a small gap\cite{PhysRevLett.116.107203,s41467-020-15594-1}, we use the complex fermion mean field theory that allows for gapless spin liquids. Furthermore, in light of the recently available INS data, we provide the dynamical spin structure factors in the relevant scattering planes for direct comparisons with the experiment. Our work offers an important theoretical basis for future experimental studies of the quantum spin liquid state that may be realized in PbCuTe$_2$O$_6$.

\begin{acknowledgments}
This work was supported by the NSERC of Canada and the Center for Quantum Materials at the University of Toronto. L.E.C.~was further supported by the Ontario Graduate Scholarship. Most of the computations were performed on the Niagara and Cedar clusters, which are hosted by SciNet and WestGrid in partnership with Compute Canada.
\end{acknowledgments}

\appendix

\section{\label{ansatzappendix}Details of the Mean Field Ansatz}
We provide details of how the mean field ansatzes are constrained by the PSG given in Table \ref{hyperkagomepsg}. We have explained in the main text that 1(a) and 1(b) are unphysical as they have vanishing bond parameters everywhere. Here, we analyze the remaining spin liquid states.

The $\mathbb{Z}_2$ spin liquids 2(a), 2(b), and 2(c) have the same $g_\mathcal{T}=i \tau_2$ but differ in $g_{C_2}=g_{S_4}$. We first show that time reversal symmetry constrains the bond parameters to be real for each of these states. Using \eqref{ansatztimereversal}, we have $u_{ij} = - (i \tau_2) u_{ij} (- i \tau_2)$, or
\begin{equation} \label{2timereversal}
\begin{pmatrix} \chi_{ij} & - \Delta_{ij}^* \\ - \Delta_{ij} & - \chi_{ij}^* \end{pmatrix} = \begin{pmatrix} \chi_{ij}^* & - \Delta_{ij} \\ - \Delta_{ij}^* & - \chi_{ij} \end{pmatrix} ,
\end{equation}
which implies $\chi_{ij} = \chi_{ij}^*$ and $\Delta_{ij}=\Delta_{ij}^*$. For the onsite term $u_{ii} \sim \lambda_i^{(1)} \tau_1 - \lambda_i^{(2)} \tau_2 + \lambda_i^{(3)} \tau_3$, one arrives at $\lambda_i^{(2)}=0$ via $u_{ii}=-(i \tau_2) u_{ii} (- i \tau_2)$. We further note from the definitions of singlet hopping \eqref{singlethopping} and singlet pairing \eqref{singletpairing} that $\chi_{ji}=\chi_{ij}^*$ ($=\chi_{ij}$ by realness) and $\Delta_{ji}=\Delta_{ij}$. Therefore, we have $u_{ij}=u_{ji}$.

For the 2(a) state, the gauge transformation associated with any space group operator $X$ is trivial, $G_X=1$. \eqref{ansatzsymmetry} then reads $u_{X(i)X(j)}=u_{ij}$, i.e.~the mean field ansatzes of symmetry-related bonds are equal. Since all the first (second) nearest neighbor bonds are symmetry-related, i.e.~they can be mapped to each other under space group operations, we can write $\chi_{ij}=\chi_1$ and $\Delta_{ij}=\Delta_1$ ($\chi_{ij}=\chi_2$ and $\Delta_{ij}=\Delta_2$) for all first (second) nearest neighbors $i$ and $j$. Similarly, all sites are symmetry-related, $u_{X(i)X(i)}=u_{ii}$ allows us to write the on-site terms as $\lambda_i^{(3)}=\lambda^{(3)}$ and $\lambda_i^{(1)}=\lambda^{(1)}$ for all sites $i$. 2(a) is known as the uniform ansatz.

Before discussing 2(b) and 2(c), we introduce a notation to describe the first and second nearest neighbor pairs. We write the site $(x,y,z;s)$ simply as $s$; if it is connected to a site $(x,y,z;s')$ within the same unit cell, we say that $s$ is connected to $s'$; if it is connected to a site in a different unit cell, e.g.~$(x+1,y,z,s')$, we say that $s$ is connected to $s'+\hat{\mathbf{x}}$. Using this notation, we list in Table \ref{nearestneighbor} the first and second nearest neighbors of all the 12 sublattices of a unit cell. The same notation is also used in Fig.~\ref{lattice}.

\begin{table}
\caption{\label{nearestneighbor} The first and second nearest neighbors of the sublattice $s$ in some unit cell $(x,y,z)$. We use a simplified notation as discussed in the text. Each site has two first nearest neighbors and four second nearest neighbors. See also Fig.~\ref{lattice}.}
\begin{ruledtabular}
\begin{tabular}{ccc}
$s$ & $1^\mathrm{st}$ n.~n. & $2^\mathrm{nd}$ n.~n. \\ \hline
$1$ & $7,9$ & $2-\hat{\mathbf{z}},3-\hat{\mathbf{z}}, 6, 12$ \\
$2$ & $5-\hat{\mathbf{x}},11+\hat{\mathbf{z}}$ & $1+\hat{\mathbf{z}},3,9,10$ \\
$3$ & $8+\hat{\mathbf{z}},12+\hat{\mathbf{y}}+\hat{\mathbf{z}}$ & $1+\hat{\mathbf{z}},2,4,5$ \\
$4$ & $6+\hat{\mathbf{y}},10+\hat{\mathbf{x}}+\hat{\mathbf{y}}$ & $8,9+\hat{\mathbf{y}},3,5$ \\
$5$ & $2+\hat{\mathbf{x}},11+\hat{\mathbf{x}}+\hat{\mathbf{z}}$ & $6,7+\hat{\mathbf{x}},3,4$ \\
$6$ & $4-\hat{\mathbf{y}},10+\hat{\mathbf{x}}$ & $5,7+\hat{\mathbf{x}},1,12$ \\
$7$ & $1,9$ & $5-\hat{\mathbf{x}},6-\hat{\mathbf{x}},8,11$ \\
$8$ & $3-\hat{\mathbf{z}},12+\hat{\mathbf{y}}$ & $4,9+\hat{\mathbf{y}},7,11$ \\
$9$ & $1,7$ & $4-\hat{\mathbf{y}},8-\hat{\mathbf{y}},2,10$ \\
$10$ & $4-\hat{\mathbf{x}}-\hat{\mathbf{y}},6-\hat{\mathbf{x}}$ & $11-\hat{\mathbf{y}}+\hat{\mathbf{z}},12-\hat{\mathbf{x}}+\hat{\mathbf{z}},2,9$ \\
$11$ & $2-\hat{\mathbf{z}},5-\hat{\mathbf{x}}-\hat{\mathbf{z}}$ & $10+\hat{\mathbf{y}}-\hat{\mathbf{z}},12-\hat{\mathbf{x}}+\hat{\mathbf{y}},7,8$ \\
$12$ & $3-\hat{\mathbf{y}}-\hat{\mathbf{z}},8-\hat{\mathbf{y}}$ & $10+\hat{\mathbf{x}}-\hat{\mathbf{z}},11+\hat{\mathbf{x}}-\hat{\mathbf{y}},1,6$
\end{tabular}
\end{ruledtabular}
\end{table}

Each site has 2 first nearest neighbors and 4 second nearest neighbors. Therefore, each unit cell has 12 first nearest neighbor bonds and 24 second nearest neighbor bonds. From Table \ref{nearestneighbor}, and using the fact that translational symmetries are realized trivially ($G_{T_i}=1$), we note that the first and second nearest neighbor bonds are uniquely defined by the sublattice indices. For example, we can talk about the bond formed by $2$ and $5$, which without ambiguity refers to the first nearest neighbor bond formed by $2$ and $5-\hat{\mathbf{x}}$ (or $2+\hat{\mathbf{x}}$ and $5$, which only differs by a translation). This allows us to introduce the shorthand notation $u_{s,s'}$ for the mean field ansatz $u_{(x,y,z,s),(x',y',z',s')}$. Such a nice property will no longer hold when we consider the third nearest neighbors and beyond.

\begin{table}
\caption{\label{rotationscrew} The actions of the space group operators $C_2$, $C_3$, and $S_4$, defined in \eqref{C3operation}-\eqref{S4operation}, on a site $(x,y,z;s)$, where $(x,y,z) \in \mathbb{Z} \times \mathbb{Z} \times \mathbb{Z}$ labels the unit cell and $s \in \lbrace 1, \ldots, 12 \rbrace$ labels the sublattice. This table is adapted from Table I in Ref.~\onlinecite{PhysRevB.95.054404}.}
\begin{ruledtabular}
\begin{tabular}{cccc}
$s$ & $C_2$ & $C_3$ & $S_4$ \\ \hline
1 & $(-x,-z,-y;7)$ & $(z,x,y;7)$ & $(x,-z,y+1;12)$ \\
2 & $(-x,-z-1,-y;8)$ & $(z,x,y;6)$ & $(x,-z-1,y+1;1)$ \\
3 & $(-x,-z-1,-y;11)$ & $(z,x,y;5)$ & $(x,-z-1,y+1;6)$ \\
4 & $(-x,-z,-y-1;10)$ & $(z,x,y;3)$ & $(x,-z-1,y+1;5)$ \\
5 & $(-x-1,-z,-y;12)$ & $(z,x,y;4)$ & $(x+1,-z-1,y+1;7)$ \\
6 & $(-x-1,-z,-y;6)$ & $(z,x,y;8)$ & $(x+1,-z-1,y+1;11)$ \\
7 & $(-x,-z,-y;1)$ & $(z,x,y;9)$ & $(x,-z-1,y+1;8)$ \\
8 & $(-x,-z,-y-1;2)$ & $(z,x,y;2)$ & $(x,-z-1,y+1;4)$ \\
9 & $(-x,-z,-y;9)$ & $(z,x,y;1)$ & $(x,-z-1,y;3)$ \\
10 & $(-x,-z-1,-y;4)$ & $(z,x,y;12)$ & $(x,-z-1,y;2)$ \\
11 & $(-x,-z,-y-1;3)$ & $(z,x,y;10)$ & $(x,-z,y+1;9)$ \\
12 & $(-x-1,-z,-y;5)$ & $(z,x,y;11)$ & $(x+1,-z,y;10)$ \\
\end{tabular}
\end{ruledtabular}
\end{table}

With this shorthand notation, \eqref{ansatzsymmetry} reads 
\begin{equation} \label{ansatzsymmetrysimple}
u_{X(s),X(s')}=g_X u_{s,s'} g_X^\dagger ,
\end{equation}
where we consider $X=C_2,C_3,S_4$ according to Table \ref{rotationscrew}. We now show that the 2(b) state cannot have finite bond parameters for first nearest neighbors. Applying $X=C_2$ to the pair $s=1$ and $s'=7$ in \eqref{ansatzsymmetrysimple}, we obtain $u_{7,1} = (i \tau_2) u_{1,7} (-i \tau_2) = -u_{1,7}$. But $u_{ij}=u_{ji}$ for all $i$ and $j$, which implies $u_{1,7}=0$. Since all first nearest neighbor bonds are symmetry-related, we have $u_{ij}=u_{1,7}=0$ for any pair of first nearest neighbors $i$ and $j$. This shows that 2(b) is unphysical.

For 2(c), we first write $u_{1,7} \sim \chi_1 \tau_3 - \Delta_1 \tau_1$ and $u_{5,4} \sim \chi_2 \tau_3 - \Delta_2 \tau_1$ on the corresponding first and second nearest neighbor bonds. By $C_2$, we have $u_{7,1} = (i \tau_3) u_{1,7} (-i \tau_3) \sim \chi_1 \tau_3 + \Delta_1 \tau_1$. $u_{7,1}=u_{1,7}$ then implies $\Delta_1=0$, so $u_{1,7} \sim \chi_1 \tau_3$, which is invariant under the gauge transformations $g_{C_2}=g_{S_4}=i \tau_3$. We thus have $u_{ij}=u_{1,7} \sim \chi_1 \tau_3$ for all first nearest neighbors $i$ and $j$. Similarly, one can show that the onsite term $u_{ii} \sim \lambda^{(3)} \tau_3$ for any site $i$.

By $C_3$, we have $u_{5,4}=u_{4,3}=u_{3,5}$. Consecutive applications of $S_4$ to each of these yields $u_{7,5}=u_{4,8}=u_{5,6}=u_{8,9}=u_{6,7}=u_{9,4} \sim \chi_2 \tau_3 + \Delta_2 \tau_1$ and $u_{8,7}=u_{7,11}=u_{11,8} \sim \chi_2 \tau_3 - \Delta_2 \tau_1$. Note the opposite signs of the pairing terms for these two sets of ansatzes. By $C_2$, we have $u_{12,10}=(i \tau_3) u_{5,4} (-i \tau_3) \sim \chi_2 \tau_3 + \Delta_2 \tau_1$. By $C_3$, we have $u_{12,10}=u_{10,11}=u_{11,12}$. Consecutive applications of $S_4$ to each of these yields $u_{10,2}=u_{1,12}=u_{2,9}=u_{12,6}=u_{9,10}=u_{6,1} \sim \chi_2 \tau_3 - \Delta_2 \tau_1$ and $u_{1,3}=u_{2,1}=u_{3,2} \sim \chi_2 \tau_3 + \Delta_2 \tau_1$. By now we have obtained the mean field ansatzes of all the 24 second nearest neighbor bonds. There is no further constraint from the PSG.

For the $U(1)$ spin liquids, the mean field ansatzes contain only the hopping terms,
\begin{equation}
u_{ij} = \begin{pmatrix} \chi_{ij} & 0 \\ 0 & \chi_{ij}^* \end{pmatrix} .
\end{equation}
As discussed in the main text, we need not explicitly introduce the Lagrange multipliers as the single occupancy constraint is automatically enforced by half filling. $U1^0$ and $U1^1$ have the same $g_\mathcal{T}=i \tau_1$ but differ in $g_{C_2}=g_{S_4}$. We first show that time reversal symmetry constrains the hopping terms to be real for each of these states. Using \eqref{ansatztimereversal}, we have $u_{ij}=-(i \tau_1) u_{ij} (-i \tau_1)$, or
\begin{equation}
\begin{pmatrix} \chi_{ij} & 0 \\ 0 & - \chi_{ij}^* \end{pmatrix} = \begin{pmatrix} \chi_{ij}* & 0 \\ 0 & - \chi_{ij} \end{pmatrix}
\end{equation}
which implies $\chi_{ij}=\chi_{ij}^*$. Consequently, we have $u_{ij}=u_{ji}$ for all $i$ and $j$.

For the $U1^0$ state, the gauge transformation associated with any space group operator $X$ is trivial, $G_X=1$. Therefore, the mean field ansatzes of symmetry-related bonds are equal, i.e.~$\chi_{ij} = \chi_1$ ($\chi_{ij}=\chi_2$) for all first (second) nearest neighbor bonds. This can be compared to the 2(a) state.

For the $U1^1$ state, applying $X=C_2$ to the first nearest neighbor pair $s=1$ and $s'=7$ in \eqref{ansatzsymmetrysimple}, we obtain $u_{7,1}=(i \tau_1) u_{1,7} (-i \tau_1) = - u_{1,7}$. But $u_{ij}=u_{ji}$, which implies $u_{1,7}=0$ and thus $u_{ij}=0$ for all first nearest neighbors $i$ and $j$. This shows that $U1^1$ is unphysical, which can be compared to the $2(b)$ state.

The mean field ansatzes of the $U(1)$ spin liquids can also be obtained more directly from their descendant $\mathbb{Z}_2$ spin liquids\cite{PhysRevB.95.054404}. Turning off all the pairing terms in 2(a) or 2(c), we get $U1^0$. Turning off all the hopping terms in 2(b) or 2(c), we get $U1^1$. Therefore, $U1^1$ has vanishing bond parameters for first nearest neighbors, which makes it unphysical.

\section{\label{hamiltonianappendix}Structure of the Hamiltonian}

After the Fourier transform \eqref{fourier}, the Hamiltonian (up to some constant) can be written as $H = \sum_\mathbf{k} \Psi_\mathbf{k}^\dagger \mathrm{D}_\mathbf{k} \Psi_\mathbf{k}$.
 
For a $\mathbb{Z}_2$ spin liquid with singlet hopping and pairing channels, we can use the basis $\Psi_\mathbf{k}=( f_{\mathbf{k} \uparrow}, f_{-\mathbf{k} \downarrow}^\dagger )$, where we have suppressed the sublattice index for brevity. The matrix $\mathrm{D}_\mathbf{k}$ has the form of a BdG Hamiltonian. Therefore, its eigenvalues come in positive-negative pairs, which we denote, just for convenience, as $\omega_{\mathbf{k} \uparrow}$ and $-\omega_{-\mathbf{k} \downarrow}$ respectively. The Hamiltonian is diagonal in the Bogoliubov basis,
\begin{equation} \label{diagonalhamiltonian}
\begin{aligned}[b]
H &= \sum_\mathbf{k} \left( \omega_{\mathbf{k} \uparrow} \gamma_{\mathbf{k} \uparrow}^\dagger \gamma_{\mathbf{k} \uparrow} - \omega_{-\mathbf{k} \downarrow} \gamma_{-\mathbf{k} \downarrow} \gamma_{-\mathbf{k} \downarrow}^\dagger \right) \\
&=  \sum_\mathbf{k} \left( \omega_{\mathbf{k} \uparrow} \gamma_{\mathbf{k} \uparrow}^\dagger \gamma_{\mathbf{k} \uparrow} + \omega_{-\mathbf{k} \downarrow} \gamma_{-\mathbf{k} \downarrow}^\dagger \gamma_{-\mathbf{k} \downarrow} \right) - \sum_\mathbf{k} \omega_{-\mathbf{k} \downarrow} .
\end{aligned}
\end{equation}
The construction above allows us to easily interpret the Bogoliubov quasiparticles as excitations that carry non-negative energies $\omega_{\mathbf{k} \uparrow},\omega_{-\mathbf{k} \downarrow} \geq 0$. In Figs.~\ref{2aJ1J2dispersion} and \ref{2cJ1J2dispersion}, we plot at each momentum $\mathbf{k}$ the eigenvalues of $\mathrm{D}_\mathbf{k}$, following the usual practice in the literature. Readers should keep in mind that negative eigenvalues are really $- \omega_{-\mathbf{k} \downarrow}$; there is no negative energy.

On the other hand, for a $U(1)$ spin liquid, we can use the basis $\Psi_\mathbf{k}=(f_{\mathbf{k} \uparrow}, f_{\mathbf{k} \downarrow})$. The matrix $\mathrm{D}_\mathbf{k}$ is simply a Hamiltonian of free fermions. With only singlet hopping channels, the $\uparrow$ and $\downarrow$ sectors are decoupled and equal to each other, so that $\mathrm{D}_\mathbf{k}$ is block diagonal. Unlike the previous case, we directly interpret the eigenvalues of $\mathrm{D}_\mathbf{k}$, which we denote by $\omega_{\mathbf{k} \alpha}$, as energies, without worrying them being negative. Each $\omega_{\mathbf{k} \alpha}$ is doubly degenerate due to the spin degeneracy. The Hamiltonian can be written as $H = \sum_{\mathbf{k}\alpha} \omega_{\mathbf{k} \alpha} \gamma_{\mathbf{k} \alpha}^\dagger \gamma_{\mathbf{k} \alpha}$. In the ground state, the lower half of the energy eigenstates are filled, while the upper half are empty, in order to satisfy the single occupancy constraint. We can define the Fermi level $\varepsilon_\mathrm{F}$ that separates the filled and empty states. An excitation corresponds to creating (removing) a quasiparticle above (below) the Fermi level.

\section{\label{dynamicappendix}Calculation of the Dynamical Spin Structure Factor}

To evaluate the dynamical spin structure factor \eqref{dynamic}, we write the time evolved spin operator as $\mathbf{S}_i (t) = e^{iHt} \mathbf{S}_i (0) e^{-iHt}$, and represent $\mathbf{S}_i (0)$ using complex fermions as in \eqref{partonrepresentation}. $\mathbf{S}_i (t) \cdot \mathbf{S}_j (0)$ becomes a summation of several terms that are quartic in the fermions $e^{iHt} f_{i \alpha_1}^\dagger f_{i \alpha_2} e^{-iHt} f_{j \alpha_3}^\dagger f_{j \alpha_4}$, where $\alpha_i \in \lbrace \uparrow, \downarrow \rbrace$. For concreteness, we demonstrate how to evaluate
\begin{equation} \label{dynamicexample}
\begin{aligned}[b]
\int \mathrm{d} t \, e^{iEt} \sum_{ij} \sum_{m,n=1}^\mathcal{N} & \left \langle e^{iHt} f_{i, m, \uparrow}^\dagger f_{i, m \downarrow} e^{-iHt} f_{j, n, \downarrow}^\dagger f_{j, n, \uparrow} \right \rangle \\
&\times e^{i \mathbf{k} \cdot (\mathbf{R}_j + \mathbf{d}_n - \mathbf{R}_i - \mathbf{d}_m)} ,
\end{aligned}
\end{equation}
a term that is contributed by $S_i^x (t) S_j^x (0)$ and $S_i^y (t) S_j^y (0)$, for a $\mathbb{Z}_2$ spin liquid. We have split the site index into the unit cell part $i$ ($j$) and the sublattice part $m$ ($n$). $\mathcal{N}=12$ is the total number of sublattices per unit cell.

We apply the Fourier transform \eqref{fourier} to each of the four fermion operators,
\begin{equation}
\begin{aligned}[b]
& \left \langle e^{iHt} f_{i, m, \uparrow}^\dagger f_{i, m \downarrow} e^{-iHt} f_{j, n, \downarrow}^\dagger f_{j, n, \uparrow} \right \rangle \\
&= \sum_{\mathbf{k}_1 \mathbf{k}_2 \mathbf{k}_3
 \mathbf{k}_4} \left \langle e^{iHt} f_{\mathbf{k}_1,m,\uparrow}^\dagger f_{-\mathbf{k}_2,m,\downarrow} e^{-iHt} f_{-\mathbf{k}_3,n,\downarrow}^\dagger f_{\mathbf{k}_4,n,\uparrow} \right \rangle \\
& \times \frac{1}{N^2} e^{-i (\mathbf{k}_1 + \mathbf{k}_2) \cdot \mathbf{R}_i} e^{i (\mathbf{k}_3+\mathbf{k}_4) \cdot \mathbf{R}_j}
\end{aligned}
\end{equation}
Let $\mathrm{U}_\mathbf{k}$ be the unitary matrix that diagonalize $\mathrm{D}_\mathbf{k}$, such that $\mathrm{U}_\mathbf{k}^\dagger \mathrm{D}_\mathbf{k} \mathrm{U}_\mathbf{k} = \mathrm{diag}(\omega_{\mathbf{k},1,\uparrow}, \ldots, \omega_{\mathbf{k},\mathcal{N},\uparrow}, - \omega_{-\mathbf{k},1,\downarrow}, \ldots, - \omega_{-\mathbf{k},\mathcal{N},\downarrow})$. The Bogoliubov quasiparticles $\gamma$ are defined via
\begin{equation*}
\begin{aligned}
& f_{\mathbf{k},s,\uparrow} = \sum_{a=1}^\mathcal{N} \left[ \mathrm{U}_\mathbf{k} (s,a) \gamma_{\mathbf{k},a,\uparrow} + \mathrm{U}_\mathbf{k} (s,a+\mathcal{N}) \gamma_{-\mathbf{k},a,\downarrow}^\dagger \right] , \\
& f_{-\mathbf{k},s,\downarrow}^\dagger = \sum_{a=1}^\mathcal{N} \left[ \mathrm{U}_\mathbf{k} (s+\mathcal{N},a) \gamma_{\mathbf{k},a,\uparrow} + \mathrm{U}_\mathbf{k} (s+\mathcal{N},a+\mathcal{N}) \gamma_{-\mathbf{k},a,\downarrow}^\dagger \right] .
\end{aligned}
\end{equation*}
For a $\mathbb{Z}_2$ spin liquid, since the ground state $\rvert 0 \rangle$ has no Bogoliubov quasiparticle, finite contribution comes from creating two Bogoliubov quasiparticles at time $0$ and then annihilating them at time $t$. Therefore,
\begin{equation} \label{twospinonstate}
\begin{aligned}[b]
& e^{-iHt} f_{-\mathbf{k}_3,n,\downarrow}^\dagger f_{\mathbf{k}_4,n,\uparrow} \rvert 0 \rangle = \sum_{a,b=1}^\mathcal{N} e^{- i (\omega_{-\mathbf{k}_3,b,\downarrow}+\omega_{-\mathbf{k}_4,a,\downarrow}) t} \\ &\times \mathrm{U}_{\mathbf{k}_3} (n+\mathcal{N},b+\mathcal{N}) \mathrm{U}_{\mathbf{k}_4} (n,a+\mathcal{N}) \gamma_{-\mathbf{k}_3,b,\downarrow}^\dagger \gamma_{-\mathbf{k}_4,a,\downarrow}^\dagger \rvert 0 \rangle
\end{aligned}
\end{equation}
We have neglected the ground state energy in the exponent as it will be cancelled anyway by the factor of $e^{iHt}$ on the left. By Pauli exclusion principle, $(\mathbf{k}_3,b) \neq (\mathbf{k}_4,a)$. Then, $f_{\mathbf{k}_1,m,\uparrow}^\dagger f_{\mathbf{k}_2,m,\downarrow}$ should annihilate $\gamma_{-\mathbf{k}_3,b,\downarrow}^\dagger \gamma_{-\mathbf{k}_4,a,\uparrow}^\dagger$. This can happen either when $\mathbf{k}_2=\mathbf{k}_3$, $\mathbf{k}_1=\mathbf{k}_4$ or when $\mathbf{k}_2=\mathbf{k}_4$, $\mathbf{k}_1=\mathbf{k}_3$.

Summing over $\mathbf{R}_j$ gives a delta function $N \delta (\mathbf{k}+\mathbf{k}_3+\mathbf{k}_4)$. Then, summing over $\mathbf{R}_i$ simply gives a factor of $N$. The time integral gives a delta function $\delta (E-\omega_{-\mathbf{k}_3,b,\downarrow}-\omega_{-\mathbf{k}_4,a,\downarrow})$. Combining all these, the final expression of \eqref{dynamicexample} is
\begin{widetext}
\begin{equation}
\begin{aligned}[b]
& \sum_{\mathbf{k}_3 \mathbf{k}_4} \sum_{mn} \sum_{a b} \delta (\mathbf{k}+\mathbf{k}_3+\mathbf{k}_4) \delta (E-\omega_{-\mathbf{k}_3,b,\downarrow}-\omega_{-\mathbf{k}_4,a,\downarrow}) e^{i \mathbf{k} \cdot (\mathbf{d}_n-\mathbf{d}_m)} \\
& \times \left[ \mathrm{U}_{\mathbf{k}_4}^* (m,a+\mathcal{N}) \mathrm{U}_{\mathbf{k}_3}^* (m+\mathcal{N},b+\mathcal{N}) - \mathrm{U}_{\mathbf{k}_3}^* (m,b+\mathcal{N}) \mathrm{U}_{\mathbf{k}_4}^* (m+\mathcal{N},a+\mathcal{N}) \right] \mathrm{U}_{\mathbf{k}_3} (n+\mathcal{N},b+\mathcal{N}) \mathrm{U}_{\mathbf{k}_4} (n, a+\mathcal{N}) .
\end{aligned}
\end{equation}
\end{widetext}
with $(\mathbf{k}_3,b) \neq (\mathbf{k}_4,a)$ as required by Pauli exclusion principle. The minus sign in the square bracket is due to the fermionic anticommutation relation.

The calculation of the dynamical spin structure factor for a $U(1)$ spin liquid proceeds along similar lines. The main differences are that (i) the quasiparticles do not mix the original fermionic creation and annihilation operators, i.e.~$\gamma$ ($\gamma^\dagger$) is a linear combination of $f$ ($f^\dagger$) only, and (ii) the ground state corresponds to filling the lower half of the energy eigenstates. Finite contribution to the structure factor comes from the following consideration. At time $0$, a pair of excitations are created by removing a quasiparticle from one of the filled states, say $\lvert \mathbf{k}_4, a \rangle$, \textit{and} adding a quasiparticle in one of the empty states, say $\lvert \mathbf{k}_3, b \rangle$. The energy required to do so is $-(\omega_{\mathbf{k}_4,a}-\varepsilon_\mathrm{F})+(\omega_{\mathbf{k}_3,b}-\varepsilon_\mathrm{F})=\omega_{\mathbf{k}_3,b}-\omega_{\mathbf{k}_4,a}$, which is always non-negative. This pair of excitations are then annihilated at time $t$. The time integral gives a delta function $\delta (E - \omega_{\mathbf{k}_3,b}+\omega_{\mathbf{k}_4,a})$.

When we have to perform a summation of momenta over the first Brillouin zone, e.g.~solving the self consistent equations or calculating the dynamical spin structure factor, we divide the first Brillouin zone evenly such that it contains $L \times L \times L$ $\mathbf{k}$ points. We choose $L$ to be at least $40$.

\bibliography{reference210305}

\begin{thebibliography}{40}%
\makeatletter
\providecommand \@ifxundefined [1]{%
 \@ifx{#1\undefined}
}%
\providecommand \@ifnum [1]{%
 \ifnum #1\expandafter \@firstoftwo
 \else \expandafter \@secondoftwo
 \fi
}%
\providecommand \@ifx [1]{%
 \ifx #1\expandafter \@firstoftwo
 \else \expandafter \@secondoftwo
 \fi
}%
\providecommand \natexlab [1]{#1}%
\providecommand \enquote  [1]{``#1''}%
\providecommand \bibnamefont  [1]{#1}%
\providecommand \bibfnamefont [1]{#1}%
\providecommand \citenamefont [1]{#1}%
\providecommand \href@noop [0]{\@secondoftwo}%
\providecommand \href [0]{\begingroup \@sanitize@url \@href}%
\providecommand \@href[1]{\@@startlink{#1}\@@href}%
\providecommand \@@href[1]{\endgroup#1\@@endlink}%
\providecommand \@sanitize@url [0]{\catcode `\\12\catcode `\$12\catcode
  `\&12\catcode `\#12\catcode `\^12\catcode `\_12\catcode `\%12\relax}%
\providecommand \@@startlink[1]{}%
\providecommand \@@endlink[0]{}%
\providecommand \url  [0]{\begingroup\@sanitize@url \@url }%
\providecommand \@url [1]{\endgroup\@href {#1}{\urlprefix }}%
\providecommand \urlprefix  [0]{URL }%
\providecommand \Eprint [0]{\href }%
\providecommand \doibase [0]{http://dx.doi.org/}%
\providecommand \selectlanguage [0]{\@gobble}%
\providecommand \bibinfo  [0]{\@secondoftwo}%
\providecommand \bibfield  [0]{\@secondoftwo}%
\providecommand \translation [1]{[#1]}%
\providecommand \BibitemOpen [0]{}%
\providecommand \bibitemStop [0]{}%
\providecommand \bibitemNoStop [0]{.\EOS\space}%
\providecommand \EOS [0]{\spacefactor3000\relax}%
\providecommand \BibitemShut  [1]{\csname bibitem#1\endcsname}%
\let\auto@bib@innerbib\@empty
\bibitem [{\citenamefont {Balents}(2010)}]{nature08917}%
  \BibitemOpen
  \bibfield  {author} {\bibinfo {author} {\bibfnamefont {L.}~\bibnamefont
  {Balents}},\ }\href {\doibase 10.1038/nature08917} {\bibfield  {journal}
  {\bibinfo  {journal} {Nature}\ }\textbf {\bibinfo {volume} {464}},\ \bibinfo
  {pages} {199} (\bibinfo {year} {2010})}\BibitemShut {NoStop}%
\bibitem [{\citenamefont {Savary}\ and\ \citenamefont
  {Balents}(2016)}]{Savary_2016}%
  \BibitemOpen
  \bibfield  {author} {\bibinfo {author} {\bibfnamefont {L.}~\bibnamefont
  {Savary}}\ and\ \bibinfo {author} {\bibfnamefont {L.}~\bibnamefont
  {Balents}},\ }\href {\doibase 10.1088/0034-4885/80/1/016502} {\bibfield
  {journal} {\bibinfo  {journal} {Reports on Progress in Physics}\ }\textbf
  {\bibinfo {volume} {80}},\ \bibinfo {pages} {016502} (\bibinfo {year}
  {2016})}\BibitemShut {NoStop}%
\bibitem [{\citenamefont {Zhou}\ \emph {et~al.}(2017)\citenamefont {Zhou},
  \citenamefont {Kanoda},\ and\ \citenamefont {Ng}}]{RevModPhys.89.025003}%
  \BibitemOpen
  \bibfield  {author} {\bibinfo {author} {\bibfnamefont {Y.}~\bibnamefont
  {Zhou}}, \bibinfo {author} {\bibfnamefont {K.}~\bibnamefont {Kanoda}}, \ and\
  \bibinfo {author} {\bibfnamefont {T.-K.}\ \bibnamefont {Ng}},\ }\href
  {\doibase 10.1103/RevModPhys.89.025003} {\bibfield  {journal} {\bibinfo
  {journal} {Rev. Mod. Phys.}\ }\textbf {\bibinfo {volume} {89}},\ \bibinfo
  {pages} {025003} (\bibinfo {year} {2017})}\BibitemShut {NoStop}%
\bibitem [{\citenamefont {Ramirez}(1994)}]{annurev.ms.24.080194.002321}%
  \BibitemOpen
  \bibfield  {author} {\bibinfo {author} {\bibfnamefont {A.~P.}\ \bibnamefont
  {Ramirez}},\ }\href {\doibase 10.1146/annurev.ms.24.080194.002321} {\bibfield
   {journal} {\bibinfo  {journal} {Annu. Rev. Mater. Sci.}\ }\textbf {\bibinfo
  {volume} {24}},\ \bibinfo {pages} {453} (\bibinfo {year} {1994})}\BibitemShut
  {NoStop}%
\bibitem [{\citenamefont {Shimizu}\ \emph {et~al.}(2003)\citenamefont
  {Shimizu}, \citenamefont {Miyagawa}, \citenamefont {Kanoda}, \citenamefont
  {Maesato},\ and\ \citenamefont {Saito}}]{PhysRevLett.91.107001}%
  \BibitemOpen
  \bibfield  {author} {\bibinfo {author} {\bibfnamefont {Y.}~\bibnamefont
  {Shimizu}}, \bibinfo {author} {\bibfnamefont {K.}~\bibnamefont {Miyagawa}},
  \bibinfo {author} {\bibfnamefont {K.}~\bibnamefont {Kanoda}}, \bibinfo
  {author} {\bibfnamefont {M.}~\bibnamefont {Maesato}}, \ and\ \bibinfo
  {author} {\bibfnamefont {G.}~\bibnamefont {Saito}},\ }\href {\doibase
  10.1103/PhysRevLett.91.107001} {\bibfield  {journal} {\bibinfo  {journal}
  {Phys. Rev. Lett.}\ }\textbf {\bibinfo {volume} {91}},\ \bibinfo {pages}
  {107001} (\bibinfo {year} {2003})}\BibitemShut {NoStop}%
\bibitem [{\citenamefont {Shimizu}\ \emph {et~al.}(2006)\citenamefont
  {Shimizu}, \citenamefont {Miyagawa}, \citenamefont {Kanoda}, \citenamefont
  {Maesato},\ and\ \citenamefont {Saito}}]{PhysRevB.73.140407}%
  \BibitemOpen
  \bibfield  {author} {\bibinfo {author} {\bibfnamefont {Y.}~\bibnamefont
  {Shimizu}}, \bibinfo {author} {\bibfnamefont {K.}~\bibnamefont {Miyagawa}},
  \bibinfo {author} {\bibfnamefont {K.}~\bibnamefont {Kanoda}}, \bibinfo
  {author} {\bibfnamefont {M.}~\bibnamefont {Maesato}}, \ and\ \bibinfo
  {author} {\bibfnamefont {G.}~\bibnamefont {Saito}},\ }\href {\doibase
  10.1103/PhysRevB.73.140407} {\bibfield  {journal} {\bibinfo  {journal} {Phys.
  Rev. B}\ }\textbf {\bibinfo {volume} {73}},\ \bibinfo {pages} {140407}
  (\bibinfo {year} {2006})}\BibitemShut {NoStop}%
\bibitem [{\citenamefont {Yamashita}\ \emph {et~al.}(2008)\citenamefont
  {Yamashita}, \citenamefont {Nakazawa}, \citenamefont {Oguni}, \citenamefont
  {Oshima}, \citenamefont {Nojiri}, \citenamefont {Shimizu}, \citenamefont
  {Miyagawa},\ and\ \citenamefont {Kanoda}}]{nphys942}%
  \BibitemOpen
  \bibfield  {author} {\bibinfo {author} {\bibfnamefont {S.}~\bibnamefont
  {Yamashita}}, \bibinfo {author} {\bibfnamefont {Y.}~\bibnamefont {Nakazawa}},
  \bibinfo {author} {\bibfnamefont {M.}~\bibnamefont {Oguni}}, \bibinfo
  {author} {\bibfnamefont {Y.}~\bibnamefont {Oshima}}, \bibinfo {author}
  {\bibfnamefont {H.}~\bibnamefont {Nojiri}}, \bibinfo {author} {\bibfnamefont
  {Y.}~\bibnamefont {Shimizu}}, \bibinfo {author} {\bibfnamefont
  {K.}~\bibnamefont {Miyagawa}}, \ and\ \bibinfo {author} {\bibfnamefont
  {K.}~\bibnamefont {Kanoda}},\ }\href {\doibase 10.1038/nphys942} {\bibfield
  {journal} {\bibinfo  {journal} {Nature Physics}\ }\textbf {\bibinfo {volume}
  {4}},\ \bibinfo {pages} {459} (\bibinfo {year} {2008})}\BibitemShut {NoStop}%
\bibitem [{\citenamefont {Yamashita}\ \emph {et~al.}(2009)\citenamefont
  {Yamashita}, \citenamefont {Nakata}, \citenamefont {Kasahara}, \citenamefont
  {Sasaki}, \citenamefont {Yoneyama}, \citenamefont {Kobayashi}, \citenamefont
  {Fujimoto}, \citenamefont {Shibauchi},\ and\ \citenamefont
  {Matsuda}}]{nphys1134}%
  \BibitemOpen
  \bibfield  {author} {\bibinfo {author} {\bibfnamefont {M.}~\bibnamefont
  {Yamashita}}, \bibinfo {author} {\bibfnamefont {N.}~\bibnamefont {Nakata}},
  \bibinfo {author} {\bibfnamefont {Y.}~\bibnamefont {Kasahara}}, \bibinfo
  {author} {\bibfnamefont {T.}~\bibnamefont {Sasaki}}, \bibinfo {author}
  {\bibfnamefont {N.}~\bibnamefont {Yoneyama}}, \bibinfo {author}
  {\bibfnamefont {N.}~\bibnamefont {Kobayashi}}, \bibinfo {author}
  {\bibfnamefont {S.}~\bibnamefont {Fujimoto}}, \bibinfo {author}
  {\bibfnamefont {T.}~\bibnamefont {Shibauchi}}, \ and\ \bibinfo {author}
  {\bibfnamefont {Y.}~\bibnamefont {Matsuda}},\ }\href {\doibase
  10.1038/nphys1134} {\bibfield  {journal} {\bibinfo  {journal} {Nature
  Physics}\ }\textbf {\bibinfo {volume} {5}},\ \bibinfo {pages} {44} (\bibinfo
  {year} {2009})}\BibitemShut {NoStop}%
\bibitem [{\citenamefont {Mendels}\ \emph {et~al.}(2007)\citenamefont
  {Mendels}, \citenamefont {Bert}, \citenamefont {de~Vries}, \citenamefont
  {Olariu}, \citenamefont {Harrison}, \citenamefont {Duc}, \citenamefont
  {Trombe}, \citenamefont {Lord}, \citenamefont {Amato},\ and\ \citenamefont
  {Baines}}]{PhysRevLett.98.077204}%
  \BibitemOpen
  \bibfield  {author} {\bibinfo {author} {\bibfnamefont {P.}~\bibnamefont
  {Mendels}}, \bibinfo {author} {\bibfnamefont {F.}~\bibnamefont {Bert}},
  \bibinfo {author} {\bibfnamefont {M.~A.}\ \bibnamefont {de~Vries}}, \bibinfo
  {author} {\bibfnamefont {A.}~\bibnamefont {Olariu}}, \bibinfo {author}
  {\bibfnamefont {A.}~\bibnamefont {Harrison}}, \bibinfo {author}
  {\bibfnamefont {F.}~\bibnamefont {Duc}}, \bibinfo {author} {\bibfnamefont
  {J.~C.}\ \bibnamefont {Trombe}}, \bibinfo {author} {\bibfnamefont {J.~S.}\
  \bibnamefont {Lord}}, \bibinfo {author} {\bibfnamefont {A.}~\bibnamefont
  {Amato}}, \ and\ \bibinfo {author} {\bibfnamefont {C.}~\bibnamefont
  {Baines}},\ }\href {\doibase 10.1103/PhysRevLett.98.077204} {\bibfield
  {journal} {\bibinfo  {journal} {Phys. Rev. Lett.}\ }\textbf {\bibinfo
  {volume} {98}},\ \bibinfo {pages} {077204} (\bibinfo {year}
  {2007})}\BibitemShut {NoStop}%
\bibitem [{\citenamefont {Helton}\ \emph {et~al.}(2007)\citenamefont {Helton},
  \citenamefont {Matan}, \citenamefont {Shores}, \citenamefont {Nytko},
  \citenamefont {Bartlett}, \citenamefont {Yoshida}, \citenamefont {Takano},
  \citenamefont {Suslov}, \citenamefont {Qiu}, \citenamefont {Chung},
  \citenamefont {Nocera},\ and\ \citenamefont {Lee}}]{PhysRevLett.98.107204}%
  \BibitemOpen
  \bibfield  {author} {\bibinfo {author} {\bibfnamefont {J.~S.}\ \bibnamefont
  {Helton}}, \bibinfo {author} {\bibfnamefont {K.}~\bibnamefont {Matan}},
  \bibinfo {author} {\bibfnamefont {M.~P.}\ \bibnamefont {Shores}}, \bibinfo
  {author} {\bibfnamefont {E.~A.}\ \bibnamefont {Nytko}}, \bibinfo {author}
  {\bibfnamefont {B.~M.}\ \bibnamefont {Bartlett}}, \bibinfo {author}
  {\bibfnamefont {Y.}~\bibnamefont {Yoshida}}, \bibinfo {author} {\bibfnamefont
  {Y.}~\bibnamefont {Takano}}, \bibinfo {author} {\bibfnamefont
  {A.}~\bibnamefont {Suslov}}, \bibinfo {author} {\bibfnamefont
  {Y.}~\bibnamefont {Qiu}}, \bibinfo {author} {\bibfnamefont {J.-H.}\
  \bibnamefont {Chung}}, \bibinfo {author} {\bibfnamefont {D.~G.}\ \bibnamefont
  {Nocera}}, \ and\ \bibinfo {author} {\bibfnamefont {Y.~S.}\ \bibnamefont
  {Lee}},\ }\href {\doibase 10.1103/PhysRevLett.98.107204} {\bibfield
  {journal} {\bibinfo  {journal} {Phys. Rev. Lett.}\ }\textbf {\bibinfo
  {volume} {98}},\ \bibinfo {pages} {107204} (\bibinfo {year}
  {2007})}\BibitemShut {NoStop}%
\bibitem [{\citenamefont {Imai}\ \emph {et~al.}(2008)\citenamefont {Imai},
  \citenamefont {Nytko}, \citenamefont {Bartlett}, \citenamefont {Shores},\
  and\ \citenamefont {Nocera}}]{PhysRevLett.100.077203}%
  \BibitemOpen
  \bibfield  {author} {\bibinfo {author} {\bibfnamefont {T.}~\bibnamefont
  {Imai}}, \bibinfo {author} {\bibfnamefont {E.~A.}\ \bibnamefont {Nytko}},
  \bibinfo {author} {\bibfnamefont {B.~M.}\ \bibnamefont {Bartlett}}, \bibinfo
  {author} {\bibfnamefont {M.~P.}\ \bibnamefont {Shores}}, \ and\ \bibinfo
  {author} {\bibfnamefont {D.~G.}\ \bibnamefont {Nocera}},\ }\href {\doibase
  10.1103/PhysRevLett.100.077203} {\bibfield  {journal} {\bibinfo  {journal}
  {Phys. Rev. Lett.}\ }\textbf {\bibinfo {volume} {100}},\ \bibinfo {pages}
  {077203} (\bibinfo {year} {2008})}\BibitemShut {NoStop}%
\bibitem [{\citenamefont {Han}\ \emph {et~al.}(2012)\citenamefont {Han},
  \citenamefont {Helton}, \citenamefont {Chu}, \citenamefont {Nocera},
  \citenamefont {Rodriguez-Rivera}, \citenamefont {Broholm},\ and\
  \citenamefont {Lee}}]{nature11659}%
  \BibitemOpen
  \bibfield  {author} {\bibinfo {author} {\bibfnamefont {T.-H.}\ \bibnamefont
  {Han}}, \bibinfo {author} {\bibfnamefont {J.~S.}\ \bibnamefont {Helton}},
  \bibinfo {author} {\bibfnamefont {S.}~\bibnamefont {Chu}}, \bibinfo {author}
  {\bibfnamefont {D.~G.}\ \bibnamefont {Nocera}}, \bibinfo {author}
  {\bibfnamefont {J.~A.}\ \bibnamefont {Rodriguez-Rivera}}, \bibinfo {author}
  {\bibfnamefont {C.}~\bibnamefont {Broholm}}, \ and\ \bibinfo {author}
  {\bibfnamefont {Y.~S.}\ \bibnamefont {Lee}},\ }\href {\doibase
  10.1038/nature11659} {\bibfield  {journal} {\bibinfo  {journal} {Nature}\
  }\textbf {\bibinfo {volume} {492}},\ \bibinfo {pages} {406} (\bibinfo {year}
  {2012})}\BibitemShut {NoStop}%
\bibitem [{\citenamefont {Fu}\ \emph {et~al.}(2015)\citenamefont {Fu},
  \citenamefont {Imai}, \citenamefont {Han},\ and\ \citenamefont
  {Lee}}]{Fu655}%
  \BibitemOpen
  \bibfield  {author} {\bibinfo {author} {\bibfnamefont {M.}~\bibnamefont
  {Fu}}, \bibinfo {author} {\bibfnamefont {T.}~\bibnamefont {Imai}}, \bibinfo
  {author} {\bibfnamefont {T.-H.}\ \bibnamefont {Han}}, \ and\ \bibinfo
  {author} {\bibfnamefont {Y.~S.}\ \bibnamefont {Lee}},\ }\href {\doibase
  10.1126/science.aab2120} {\bibfield  {journal} {\bibinfo  {journal}
  {Science}\ }\textbf {\bibinfo {volume} {350}},\ \bibinfo {pages} {655}
  (\bibinfo {year} {2015})}\BibitemShut {NoStop}%
\bibitem [{\citenamefont {Okamoto}\ \emph {et~al.}(2007)\citenamefont
  {Okamoto}, \citenamefont {Nohara}, \citenamefont {Aruga-Katori},\ and\
  \citenamefont {Takagi}}]{PhysRevLett.99.137207}%
  \BibitemOpen
  \bibfield  {author} {\bibinfo {author} {\bibfnamefont {Y.}~\bibnamefont
  {Okamoto}}, \bibinfo {author} {\bibfnamefont {M.}~\bibnamefont {Nohara}},
  \bibinfo {author} {\bibfnamefont {H.}~\bibnamefont {Aruga-Katori}}, \ and\
  \bibinfo {author} {\bibfnamefont {H.}~\bibnamefont {Takagi}},\ }\href
  {\doibase 10.1103/PhysRevLett.99.137207} {\bibfield  {journal} {\bibinfo
  {journal} {Phys. Rev. Lett.}\ }\textbf {\bibinfo {volume} {99}},\ \bibinfo
  {pages} {137207} (\bibinfo {year} {2007})}\BibitemShut {NoStop}%
\bibitem [{\citenamefont {Singh}\ \emph {et~al.}(2013)\citenamefont {Singh},
  \citenamefont {Tokiwa}, \citenamefont {Dong},\ and\ \citenamefont
  {Gegenwart}}]{PhysRevB.88.220413}%
  \BibitemOpen
  \bibfield  {author} {\bibinfo {author} {\bibfnamefont {Y.}~\bibnamefont
  {Singh}}, \bibinfo {author} {\bibfnamefont {Y.}~\bibnamefont {Tokiwa}},
  \bibinfo {author} {\bibfnamefont {J.}~\bibnamefont {Dong}}, \ and\ \bibinfo
  {author} {\bibfnamefont {P.}~\bibnamefont {Gegenwart}},\ }\href {\doibase
  10.1103/PhysRevB.88.220413} {\bibfield  {journal} {\bibinfo  {journal} {Phys.
  Rev. B}\ }\textbf {\bibinfo {volume} {88}},\ \bibinfo {pages} {220413}
  (\bibinfo {year} {2013})}\BibitemShut {NoStop}%
\bibitem [{\citenamefont {Shockley}\ \emph {et~al.}(2015)\citenamefont
  {Shockley}, \citenamefont {Bert}, \citenamefont {Orain}, \citenamefont
  {Okamoto},\ and\ \citenamefont {Mendels}}]{PhysRevLett.115.047201}%
  \BibitemOpen
  \bibfield  {author} {\bibinfo {author} {\bibfnamefont {A.~C.}\ \bibnamefont
  {Shockley}}, \bibinfo {author} {\bibfnamefont {F.}~\bibnamefont {Bert}},
  \bibinfo {author} {\bibfnamefont {J.-C.}\ \bibnamefont {Orain}}, \bibinfo
  {author} {\bibfnamefont {Y.}~\bibnamefont {Okamoto}}, \ and\ \bibinfo
  {author} {\bibfnamefont {P.}~\bibnamefont {Mendels}},\ }\href {\doibase
  10.1103/PhysRevLett.115.047201} {\bibfield  {journal} {\bibinfo  {journal}
  {Phys. Rev. Lett.}\ }\textbf {\bibinfo {volume} {115}},\ \bibinfo {pages}
  {047201} (\bibinfo {year} {2015})}\BibitemShut {NoStop}%
\bibitem [{\citenamefont {Hopkinson}\ \emph {et~al.}(2007)\citenamefont
  {Hopkinson}, \citenamefont {Isakov}, \citenamefont {Kee},\ and\ \citenamefont
  {Kim}}]{PhysRevLett.99.037201}%
  \BibitemOpen
  \bibfield  {author} {\bibinfo {author} {\bibfnamefont {J.~M.}\ \bibnamefont
  {Hopkinson}}, \bibinfo {author} {\bibfnamefont {S.~V.}\ \bibnamefont
  {Isakov}}, \bibinfo {author} {\bibfnamefont {H.-Y.}\ \bibnamefont {Kee}}, \
  and\ \bibinfo {author} {\bibfnamefont {Y.~B.}\ \bibnamefont {Kim}},\ }\href
  {\doibase 10.1103/PhysRevLett.99.037201} {\bibfield  {journal} {\bibinfo
  {journal} {Phys. Rev. Lett.}\ }\textbf {\bibinfo {volume} {99}},\ \bibinfo
  {pages} {037201} (\bibinfo {year} {2007})}\BibitemShut {NoStop}%
\bibitem [{\citenamefont {Lawler}\ \emph
  {et~al.}(2008{\natexlab{a}})\citenamefont {Lawler}, \citenamefont {Kee},
  \citenamefont {Kim},\ and\ \citenamefont
  {Vishwanath}}]{PhysRevLett.100.227201}%
  \BibitemOpen
  \bibfield  {author} {\bibinfo {author} {\bibfnamefont {M.~J.}\ \bibnamefont
  {Lawler}}, \bibinfo {author} {\bibfnamefont {H.-Y.}\ \bibnamefont {Kee}},
  \bibinfo {author} {\bibfnamefont {Y.~B.}\ \bibnamefont {Kim}}, \ and\
  \bibinfo {author} {\bibfnamefont {A.}~\bibnamefont {Vishwanath}},\ }\href
  {\doibase 10.1103/PhysRevLett.100.227201} {\bibfield  {journal} {\bibinfo
  {journal} {Phys. Rev. Lett.}\ }\textbf {\bibinfo {volume} {100}},\ \bibinfo
  {pages} {227201} (\bibinfo {year} {2008}{\natexlab{a}})}\BibitemShut
  {NoStop}%
\bibitem [{\citenamefont {Chen}\ and\ \citenamefont
  {Balents}(2008)}]{PhysRevB.78.094403}%
  \BibitemOpen
  \bibfield  {author} {\bibinfo {author} {\bibfnamefont {G.}~\bibnamefont
  {Chen}}\ and\ \bibinfo {author} {\bibfnamefont {L.}~\bibnamefont {Balents}},\
  }\href {\doibase 10.1103/PhysRevB.78.094403} {\bibfield  {journal} {\bibinfo
  {journal} {Phys. Rev. B}\ }\textbf {\bibinfo {volume} {78}},\ \bibinfo
  {pages} {094403} (\bibinfo {year} {2008})}\BibitemShut {NoStop}%
\bibitem [{\citenamefont {Zhou}\ \emph {et~al.}(2008)\citenamefont {Zhou},
  \citenamefont {Lee}, \citenamefont {Ng},\ and\ \citenamefont
  {Zhang}}]{PhysRevLett.101.197201}%
  \BibitemOpen
  \bibfield  {author} {\bibinfo {author} {\bibfnamefont {Y.}~\bibnamefont
  {Zhou}}, \bibinfo {author} {\bibfnamefont {P.~A.}\ \bibnamefont {Lee}},
  \bibinfo {author} {\bibfnamefont {T.-K.}\ \bibnamefont {Ng}}, \ and\ \bibinfo
  {author} {\bibfnamefont {F.-C.}\ \bibnamefont {Zhang}},\ }\href {\doibase
  10.1103/PhysRevLett.101.197201} {\bibfield  {journal} {\bibinfo  {journal}
  {Phys. Rev. Lett.}\ }\textbf {\bibinfo {volume} {101}},\ \bibinfo {pages}
  {197201} (\bibinfo {year} {2008})}\BibitemShut {NoStop}%
\bibitem [{\citenamefont {Lawler}\ \emph
  {et~al.}(2008{\natexlab{b}})\citenamefont {Lawler}, \citenamefont
  {Paramekanti}, \citenamefont {Kim},\ and\ \citenamefont
  {Balents}}]{PhysRevLett.101.197202}%
  \BibitemOpen
  \bibfield  {author} {\bibinfo {author} {\bibfnamefont {M.~J.}\ \bibnamefont
  {Lawler}}, \bibinfo {author} {\bibfnamefont {A.}~\bibnamefont {Paramekanti}},
  \bibinfo {author} {\bibfnamefont {Y.~B.}\ \bibnamefont {Kim}}, \ and\
  \bibinfo {author} {\bibfnamefont {L.}~\bibnamefont {Balents}},\ }\href
  {\doibase 10.1103/PhysRevLett.101.197202} {\bibfield  {journal} {\bibinfo
  {journal} {Phys. Rev. Lett.}\ }\textbf {\bibinfo {volume} {101}},\ \bibinfo
  {pages} {197202} (\bibinfo {year} {2008}{\natexlab{b}})}\BibitemShut
  {NoStop}%
\bibitem [{\citenamefont {Singh}\ and\ \citenamefont
  {Oitmaa}(2012)}]{PhysRevB.85.104406}%
  \BibitemOpen
  \bibfield  {author} {\bibinfo {author} {\bibfnamefont {R.~R.~P.}\
  \bibnamefont {Singh}}\ and\ \bibinfo {author} {\bibfnamefont
  {J.}~\bibnamefont {Oitmaa}},\ }\href {\doibase 10.1103/PhysRevB.85.104406}
  {\bibfield  {journal} {\bibinfo  {journal} {Phys. Rev. B}\ }\textbf {\bibinfo
  {volume} {85}},\ \bibinfo {pages} {104406} (\bibinfo {year}
  {2012})}\BibitemShut {NoStop}%
\bibitem [{\citenamefont {Chen}\ and\ \citenamefont
  {Kim}(2013)}]{PhysRevB.87.165120}%
  \BibitemOpen
  \bibfield  {author} {\bibinfo {author} {\bibfnamefont {G.}~\bibnamefont
  {Chen}}\ and\ \bibinfo {author} {\bibfnamefont {Y.~B.}\ \bibnamefont {Kim}},\
  }\href {\doibase 10.1103/PhysRevB.87.165120} {\bibfield  {journal} {\bibinfo
  {journal} {Phys. Rev. B}\ }\textbf {\bibinfo {volume} {87}},\ \bibinfo
  {pages} {165120} (\bibinfo {year} {2013})}\BibitemShut {NoStop}%
\bibitem [{\citenamefont {Shindou}(2016)}]{PhysRevB.93.094419}%
  \BibitemOpen
  \bibfield  {author} {\bibinfo {author} {\bibfnamefont {R.}~\bibnamefont
  {Shindou}},\ }\href {\doibase 10.1103/PhysRevB.93.094419} {\bibfield
  {journal} {\bibinfo  {journal} {Phys. Rev. B}\ }\textbf {\bibinfo {volume}
  {93}},\ \bibinfo {pages} {094419} (\bibinfo {year} {2016})}\BibitemShut
  {NoStop}%
\bibitem [{\citenamefont {Mizoguchi}\ \emph {et~al.}(2016)\citenamefont
  {Mizoguchi}, \citenamefont {Hwang}, \citenamefont {Lee},\ and\ \citenamefont
  {Kim}}]{PhysRevB.94.064416}%
  \BibitemOpen
  \bibfield  {author} {\bibinfo {author} {\bibfnamefont {T.}~\bibnamefont
  {Mizoguchi}}, \bibinfo {author} {\bibfnamefont {K.}~\bibnamefont {Hwang}},
  \bibinfo {author} {\bibfnamefont {E.~K.-H.}\ \bibnamefont {Lee}}, \ and\
  \bibinfo {author} {\bibfnamefont {Y.~B.}\ \bibnamefont {Kim}},\ }\href
  {\doibase 10.1103/PhysRevB.94.064416} {\bibfield  {journal} {\bibinfo
  {journal} {Phys. Rev. B}\ }\textbf {\bibinfo {volume} {94}},\ \bibinfo
  {pages} {064416} (\bibinfo {year} {2016})}\BibitemShut {NoStop}%
\bibitem [{\citenamefont {Buessen}\ and\ \citenamefont
  {Trebst}(2016)}]{PhysRevB.94.235138}%
  \BibitemOpen
  \bibfield  {author} {\bibinfo {author} {\bibfnamefont {F.~L.}\ \bibnamefont
  {Buessen}}\ and\ \bibinfo {author} {\bibfnamefont {S.}~\bibnamefont
  {Trebst}},\ }\href {\doibase 10.1103/PhysRevB.94.235138} {\bibfield
  {journal} {\bibinfo  {journal} {Phys. Rev. B}\ }\textbf {\bibinfo {volume}
  {94}},\ \bibinfo {pages} {235138} (\bibinfo {year} {2016})}\BibitemShut
  {NoStop}%
\bibitem [{\citenamefont {Huang}\ \emph {et~al.}(2017)\citenamefont {Huang},
  \citenamefont {Kim},\ and\ \citenamefont {Lu}}]{PhysRevB.95.054404}%
  \BibitemOpen
  \bibfield  {author} {\bibinfo {author} {\bibfnamefont {B.}~\bibnamefont
  {Huang}}, \bibinfo {author} {\bibfnamefont {Y.~B.}\ \bibnamefont {Kim}}, \
  and\ \bibinfo {author} {\bibfnamefont {Y.-M.}\ \bibnamefont {Lu}},\ }\href
  {\doibase 10.1103/PhysRevB.95.054404} {\bibfield  {journal} {\bibinfo
  {journal} {Phys. Rev. B}\ }\textbf {\bibinfo {volume} {95}},\ \bibinfo
  {pages} {054404} (\bibinfo {year} {2017})}\BibitemShut {NoStop}%
\bibitem [{\citenamefont {Koteswararao}\ \emph {et~al.}(2014)\citenamefont
  {Koteswararao}, \citenamefont {Kumar}, \citenamefont {Khuntia}, \citenamefont
  {Bhowal}, \citenamefont {Panda}, \citenamefont {Rahman}, \citenamefont
  {Mahajan}, \citenamefont {Dasgupta}, \citenamefont {Baenitz}, \citenamefont
  {Kim},\ and\ \citenamefont {Chou}}]{PhysRevB.90.035141}%
  \BibitemOpen
  \bibfield  {author} {\bibinfo {author} {\bibfnamefont {B.}~\bibnamefont
  {Koteswararao}}, \bibinfo {author} {\bibfnamefont {R.}~\bibnamefont {Kumar}},
  \bibinfo {author} {\bibfnamefont {P.}~\bibnamefont {Khuntia}}, \bibinfo
  {author} {\bibfnamefont {S.}~\bibnamefont {Bhowal}}, \bibinfo {author}
  {\bibfnamefont {S.~K.}\ \bibnamefont {Panda}}, \bibinfo {author}
  {\bibfnamefont {M.~R.}\ \bibnamefont {Rahman}}, \bibinfo {author}
  {\bibfnamefont {A.~V.}\ \bibnamefont {Mahajan}}, \bibinfo {author}
  {\bibfnamefont {I.}~\bibnamefont {Dasgupta}}, \bibinfo {author}
  {\bibfnamefont {M.}~\bibnamefont {Baenitz}}, \bibinfo {author} {\bibfnamefont
  {K.~H.}\ \bibnamefont {Kim}}, \ and\ \bibinfo {author} {\bibfnamefont
  {F.~C.}\ \bibnamefont {Chou}},\ }\href {\doibase 10.1103/PhysRevB.90.035141}
  {\bibfield  {journal} {\bibinfo  {journal} {Phys. Rev. B}\ }\textbf {\bibinfo
  {volume} {90}},\ \bibinfo {pages} {035141} (\bibinfo {year}
  {2014})}\BibitemShut {NoStop}%
\bibitem [{\citenamefont {Khuntia}\ \emph {et~al.}(2016)\citenamefont
  {Khuntia}, \citenamefont {Bert}, \citenamefont {Mendels}, \citenamefont
  {Koteswararao}, \citenamefont {Mahajan}, \citenamefont {Baenitz},
  \citenamefont {Chou}, \citenamefont {Baines}, \citenamefont {Amato},\ and\
  \citenamefont {Furukawa}}]{PhysRevLett.116.107203}%
  \BibitemOpen
  \bibfield  {author} {\bibinfo {author} {\bibfnamefont {P.}~\bibnamefont
  {Khuntia}}, \bibinfo {author} {\bibfnamefont {F.}~\bibnamefont {Bert}},
  \bibinfo {author} {\bibfnamefont {P.}~\bibnamefont {Mendels}}, \bibinfo
  {author} {\bibfnamefont {B.}~\bibnamefont {Koteswararao}}, \bibinfo {author}
  {\bibfnamefont {A.~V.}\ \bibnamefont {Mahajan}}, \bibinfo {author}
  {\bibfnamefont {M.}~\bibnamefont {Baenitz}}, \bibinfo {author} {\bibfnamefont
  {F.~C.}\ \bibnamefont {Chou}}, \bibinfo {author} {\bibfnamefont
  {C.}~\bibnamefont {Baines}}, \bibinfo {author} {\bibfnamefont
  {A.}~\bibnamefont {Amato}}, \ and\ \bibinfo {author} {\bibfnamefont
  {Y.}~\bibnamefont {Furukawa}},\ }\href {\doibase
  10.1103/PhysRevLett.116.107203} {\bibfield  {journal} {\bibinfo  {journal}
  {Phys. Rev. Lett.}\ }\textbf {\bibinfo {volume} {116}},\ \bibinfo {pages}
  {107203} (\bibinfo {year} {2016})}\BibitemShut {NoStop}%
\bibitem [{\citenamefont {Chillal}\ \emph {et~al.}(2020)\citenamefont
  {Chillal}, \citenamefont {Iqbal}, \citenamefont {Jeschke}, \citenamefont
  {Rodriguez-Rivera}, \citenamefont {Bewley}, \citenamefont {Manuel},
  \citenamefont {Khalyavin}, \citenamefont {Steffens}, \citenamefont {Thomale},
  \citenamefont {Islam}, \citenamefont {Reuther},\ and\ \citenamefont
  {Lake}}]{s41467-020-15594-1}%
  \BibitemOpen
  \bibfield  {author} {\bibinfo {author} {\bibfnamefont {S.}~\bibnamefont
  {Chillal}}, \bibinfo {author} {\bibfnamefont {Y.}~\bibnamefont {Iqbal}},
  \bibinfo {author} {\bibfnamefont {H.~O.}\ \bibnamefont {Jeschke}}, \bibinfo
  {author} {\bibfnamefont {J.~A.}\ \bibnamefont {Rodriguez-Rivera}}, \bibinfo
  {author} {\bibfnamefont {R.}~\bibnamefont {Bewley}}, \bibinfo {author}
  {\bibfnamefont {P.}~\bibnamefont {Manuel}}, \bibinfo {author} {\bibfnamefont
  {D.}~\bibnamefont {Khalyavin}}, \bibinfo {author} {\bibfnamefont
  {P.}~\bibnamefont {Steffens}}, \bibinfo {author} {\bibfnamefont
  {R.}~\bibnamefont {Thomale}}, \bibinfo {author} {\bibfnamefont {A.~T. M.~N.}\
  \bibnamefont {Islam}}, \bibinfo {author} {\bibfnamefont {J.}~\bibnamefont
  {Reuther}}, \ and\ \bibinfo {author} {\bibfnamefont {B.}~\bibnamefont
  {Lake}},\ }\href {\doibase 10.1038/s41467-020-15594-1} {\bibfield  {journal}
  {\bibinfo  {journal} {Nature Communications}\ }\textbf {\bibinfo {volume}
  {11}},\ \bibinfo {pages} {2348} (\bibinfo {year} {2020})}\BibitemShut
  {NoStop}%
\bibitem [{\citenamefont {Wen}(1991)}]{PhysRevB.44.2664}%
  \BibitemOpen
  \bibfield  {author} {\bibinfo {author} {\bibfnamefont {X.~G.}\ \bibnamefont
  {Wen}},\ }\href {\doibase 10.1103/PhysRevB.44.2664} {\bibfield  {journal}
  {\bibinfo  {journal} {Phys. Rev. B}\ }\textbf {\bibinfo {volume} {44}},\
  \bibinfo {pages} {2664} (\bibinfo {year} {1991})}\BibitemShut {NoStop}%
\bibitem [{\citenamefont {Mudry}\ and\ \citenamefont
  {Fradkin}(1994)}]{PhysRevB.49.5200}%
  \BibitemOpen
  \bibfield  {author} {\bibinfo {author} {\bibfnamefont {C.}~\bibnamefont
  {Mudry}}\ and\ \bibinfo {author} {\bibfnamefont {E.}~\bibnamefont
  {Fradkin}},\ }\href {\doibase 10.1103/PhysRevB.49.5200} {\bibfield  {journal}
  {\bibinfo  {journal} {Phys. Rev. B}\ }\textbf {\bibinfo {volume} {49}},\
  \bibinfo {pages} {5200} (\bibinfo {year} {1994})}\BibitemShut {NoStop}%
\bibitem [{\citenamefont {Wen}(2002)}]{PhysRevB.65.165113}%
  \BibitemOpen
  \bibfield  {author} {\bibinfo {author} {\bibfnamefont {X.-G.}\ \bibnamefont
  {Wen}},\ }\href {\doibase 10.1103/PhysRevB.65.165113} {\bibfield  {journal}
  {\bibinfo  {journal} {Phys. Rev. B}\ }\textbf {\bibinfo {volume} {65}},\
  \bibinfo {pages} {165113} (\bibinfo {year} {2002})}\BibitemShut {NoStop}%
\bibitem [{\citenamefont {Lu}\ \emph {et~al.}(2011)\citenamefont {Lu},
  \citenamefont {Ran},\ and\ \citenamefont {Lee}}]{PhysRevB.83.224413}%
  \BibitemOpen
  \bibfield  {author} {\bibinfo {author} {\bibfnamefont {Y.-M.}\ \bibnamefont
  {Lu}}, \bibinfo {author} {\bibfnamefont {Y.}~\bibnamefont {Ran}}, \ and\
  \bibinfo {author} {\bibfnamefont {P.~A.}\ \bibnamefont {Lee}},\ }\href
  {\doibase 10.1103/PhysRevB.83.224413} {\bibfield  {journal} {\bibinfo
  {journal} {Phys. Rev. B}\ }\textbf {\bibinfo {volume} {83}},\ \bibinfo
  {pages} {224413} (\bibinfo {year} {2011})}\BibitemShut {NoStop}%
\bibitem [{\citenamefont {Jin}\ and\ \citenamefont
  {Zhou}(2020)}]{PhysRevB.101.054408}%
  \BibitemOpen
  \bibfield  {author} {\bibinfo {author} {\bibfnamefont {H.-K.}\ \bibnamefont
  {Jin}}\ and\ \bibinfo {author} {\bibfnamefont {Y.}~\bibnamefont {Zhou}},\
  }\href {\doibase 10.1103/PhysRevB.101.054408} {\bibfield  {journal} {\bibinfo
   {journal} {Phys. Rev. B}\ }\textbf {\bibinfo {volume} {101}},\ \bibinfo
  {pages} {054408} (\bibinfo {year} {2020})}\BibitemShut {NoStop}%
\bibitem [{\citenamefont {Isakov}\ \emph {et~al.}(2008)\citenamefont {Isakov},
  \citenamefont {Hopkinson},\ and\ \citenamefont {Kee}}]{PhysRevB.78.014404}%
  \BibitemOpen
  \bibfield  {author} {\bibinfo {author} {\bibfnamefont {S.~V.}\ \bibnamefont
  {Isakov}}, \bibinfo {author} {\bibfnamefont {J.~M.}\ \bibnamefont
  {Hopkinson}}, \ and\ \bibinfo {author} {\bibfnamefont {H.-Y.}\ \bibnamefont
  {Kee}},\ }\href {\doibase 10.1103/PhysRevB.78.014404} {\bibfield  {journal}
  {\bibinfo  {journal} {Phys. Rev. B}\ }\textbf {\bibinfo {volume} {78}},\
  \bibinfo {pages} {014404} (\bibinfo {year} {2008})}\BibitemShut {NoStop}%
\bibitem [{\citenamefont {Affleck}\ \emph {et~al.}(1988)\citenamefont
  {Affleck}, \citenamefont {Zou}, \citenamefont {Hsu},\ and\ \citenamefont
  {Anderson}}]{PhysRevB.38.745}%
  \BibitemOpen
  \bibfield  {author} {\bibinfo {author} {\bibfnamefont {I.}~\bibnamefont
  {Affleck}}, \bibinfo {author} {\bibfnamefont {Z.}~\bibnamefont {Zou}},
  \bibinfo {author} {\bibfnamefont {T.}~\bibnamefont {Hsu}}, \ and\ \bibinfo
  {author} {\bibfnamefont {P.~W.}\ \bibnamefont {Anderson}},\ }\href {\doibase
  10.1103/PhysRevB.38.745} {\bibfield  {journal} {\bibinfo  {journal} {Phys.
  Rev. B}\ }\textbf {\bibinfo {volume} {38}},\ \bibinfo {pages} {745} (\bibinfo
  {year} {1988})}\BibitemShut {NoStop}%
\bibitem [{\citenamefont {Dagotto}\ \emph {et~al.}(1988)\citenamefont
  {Dagotto}, \citenamefont {Fradkin},\ and\ \citenamefont
  {Moreo}}]{PhysRevB.38.2926}%
  \BibitemOpen
  \bibfield  {author} {\bibinfo {author} {\bibfnamefont {E.}~\bibnamefont
  {Dagotto}}, \bibinfo {author} {\bibfnamefont {E.}~\bibnamefont {Fradkin}}, \
  and\ \bibinfo {author} {\bibfnamefont {A.}~\bibnamefont {Moreo}},\ }\href
  {\doibase 10.1103/PhysRevB.38.2926} {\bibfield  {journal} {\bibinfo
  {journal} {Phys. Rev. B}\ }\textbf {\bibinfo {volume} {38}},\ \bibinfo
  {pages} {2926} (\bibinfo {year} {1988})}\BibitemShut {NoStop}%
\bibitem [{ene()}]{energydifference}%
  \BibitemOpen
  \href@noop {} {}\bibinfo {note} {The energies per site of 2(c) and $U1^0$ are
  equal up to three significant figures, but difference can be seen if we
  include more significant figures, e.g.~$-0.096628 J_1$ for 2(c) and
  $-0.096598 J_1$ for $U1^0$. This shows that a small pairing term can slightly
  lower the energy.}\BibitemShut {Stop}%
\bibitem [{\citenamefont {Setyawan}\ and\ \citenamefont
  {Curtarolo}(2010)}]{SETYAWAN2010299}%
  \BibitemOpen
  \bibfield  {author} {\bibinfo {author} {\bibfnamefont {W.}~\bibnamefont
  {Setyawan}}\ and\ \bibinfo {author} {\bibfnamefont {S.}~\bibnamefont
  {Curtarolo}},\ }\href {\doibase
  https://doi.org/10.1016/j.commatsci.2010.05.010} {\bibfield  {journal}
  {\bibinfo  {journal} {Computational Materials Science}\ }\textbf {\bibinfo
  {volume} {49}},\ \bibinfo {pages} {299} (\bibinfo {year} {2010})}\BibitemShut
  {NoStop}%
\end{thebibliography}%

\end{document}